\documentclass[twocolumn]{aastex62}

\graphicspath{{./}{figures/}}

\usepackage{amsmath}
\usepackage{float}
\accepted{ApJ}

\shorttitle{DM halo shape: MW like galaxies}
\shortauthors{Razieh Emami et. al.}

\begin{document}

\title{DM halo morphological types of MW-like galaxies in the TNG50 simulation: Simple, Twisted, or Stretched}

\correspondingauthor{Razieh Emami}
\email{razieh.emami$_{-}$meibody@cfa.harvard.edu}

\author{Razieh Emami}
\affiliation{Center for Astrophysics $\vert$ Harvard \& Smithsonian, 60 Garden Street, Cambridge, MA 02138, USA}

\author{Shy Genel}
\affiliation{Center for Computational Astrophysics, Flatiron Institute, New York, USA}
\affiliation{Columbia Astrophysics Laboratory, Columbia University, 550 West 120th Street, New York, NY 10027, USA}

\author{Lars Hernquist}
\affiliation{Center for Astrophysics $\vert$ Harvard \& Smithsonian, 60 Garden Street, Cambridge, MA 02138, USA}

\author{Charles Alcock}
\affiliation{Center for Astrophysics $\vert$ Harvard \& Smithsonian, 60 Garden Street, Cambridge, MA 02138, USA}

\author{Sownak Bose}
\affiliation{Center for Astrophysics $\vert$ Harvard \& Smithsonian, 60 Garden Street, Cambridge, MA 02138, USA}

\author{Rainer Weinberger}
\affiliation{Center for Astrophysics $\vert$ Harvard \& Smithsonian, 60 Garden Street, Cambridge, MA 02138, USA}

\author{Mark Vogelsberger}
\affiliation{Department of Physics, Kavli Institute for Astrophysics and Space Research, Massachusetts Institute of Technology, Cambridge, MA 02139, USA}

\author{Federico Marinacci}
\affiliation{Department of Physics \& Astronomy "Augusto Righi", University of Bologna, via Gobetti 93/2, 40129 Bologna, Italy}

\author{Abraham Loeb}
\affiliation{Center for Astrophysics $\vert$ Harvard \& Smithsonian, 60 Garden Street, Cambridge, MA 02138, USA}

\author{Paul Torrey}
\affiliation{Department of Astronomy, University of Florida, 211 Bryant Space Sciences Center, Gainesville, FL 32611, USA}

\author{John C. Forbes}
\affiliation{Center for Computational Astrophysics, Flatiron Institute, New York, USA}

\begin{abstract}

We present a comprehensive analysis of the shape of dark matter(DM) halos in a sample of 25 Milky Way-like galaxies in TNG50 simulation. Using an Enclosed Volume Iterative Method(EVIM), we infer an oblate-to-triaxial shape for the DM halo with the median $T \simeq 0.24 $. We group DM halos in 3 different categories. Simple halos (32\% of population) establish principal axes whose ordering in magnitude does not change with radius
and whose orientations are almost fixed throughout the halo. Twisted halos (32\% of population), experience levels of gradual rotations throughout their radial profiles. Finally, stretched halos (36\% of population) demonstrate a stretching in their principal axes lengths where the ordering of different eigenvalues change with radius. Subsequently, the halo experiences a `rotation' of $\sim$90 \rm{deg} where the stretching occurs. Visualizing the 3D ellipsoid of each halo, for the first time, we report signs of re-orienting ellipsoid in twisted and stretched halos. 
We examine the impact of baryonic physics on DM halo shape through a comparison to dark matter only(DMO) simulations. This suggests a triaxial(prolate) halo. 
We analyze the impact of substructure on DM halo shape in both hydro and DMO simulations and confirm that their impacts are subdominant. 
We study the distribution of satellites in our sample. In simple and twisted halos, the angle of satellites' angular momentum with galaxy's angular momentum grows with radius. However, stretched halos show a flat distribution of angles. Overlaying our theoretical outcome on the observational results presented in the literature establishes a fair agreement.

\end{abstract}
 

\section{Introduction}
The morphologies of galaxies are affected by a number of dynamical mechanisms operating throughout the history of galaxy evolution. Star formation, accretion of gas, galaxy mergers and feedback from supermassive black holes (SMBH) play an essential role in shaping galaxies. Structural morphology of galaxies, in general, can be assessed from the three-dimensional shape analysis. As the 3D shape of a galaxy describes the spatial distribution of mass, it provides an exceptional description of galaxy morphology. 

Observationally, several studies have measured the shape of the dark matter (DM) halo of the Milky Way (MW) galaxy where to infer the shape of the DM halo, different tracers have been used such as the orbit of Sgr dwarf \citep{2001ApJ...551..294I,2005ApJ...619..800J}, the radial velocities \citep{2005ApJ...619..800J} or the line of sight velocities \citep{2004ApJ...610L..97H} of M giant candidates for modeling Sgr dwarf debris, kinematic of K dwarf \citep{2012MNRAS.425.1445G}, tidal stream from Palomar5 (Pal5) \citep{2015ApJ...803...80K} or the proper motion of globular clusters \citep{2019A&A...621A..56P}. As pointed out below, each of these techniques gives rise to slightly different results. It is therefore beneficial to use high resolution cosmological simulations, which cover all of the aforementioned effects, to estimate the shape of halos and compare them with different observational results.

There have been significant developments in the study of galaxy morphology using high-resolution hydrodynamical simulations such as AURIGA \citep{2016MNRAS.459L..46M, 2018MNRAS.481.1726G, 2019MNRAS.488..135H}, 
EAGLE \citep{2015MNRAS.446..521S, 2015MNRAS.450.1937C, 2019MNRAS.483..744T, 2020arXiv200401914F, 2020arXiv200103178S}, FIRE \citep{2018MNRAS.481.4133G, 2018MNRAS.473.1930E,2019arXiv191100020O, 2020ApJS..246....6S} and NIHAO-UHD \citep{2018IAUS..334..209B, 2020MNRAS.491.3461B}.
In addition, there have been a number of such studies using 
the Illustris simulation \citep{2014MNRAS.444.1518V,2014Natur.509..177V, 2014MNRAS.445..175G, 2015MNRAS.452..575S} 
as well as the IllustrisTNG simulations
\citep{2018MNRAS.477.1206N, 2018MNRAS.475..648P,2018MNRAS.475..676S, 2018MNRAS.475..624N,2018MNRAS.480.5113M, 2020NatRP...2...42V, 2020MNRAS.tmp.1340M}. 
Thanks to these modern simulations, we can produce realistic galaxy populations with reliable data suitable for our theoretical investigations. The shapes of elliptical and spiral galaxies are well described with triaxial ellipsoids \citep{1970ApJ...160..831S, 1992MNRAS.258..404L} with three different axis lengths $(a \leq b \leq c)$. From these axis lengths we can read three broad geometrical types; oblate (disky) galaxies have ($c \simeq b > a$), prolate (elongated) galaxies are associated with ($c > b \simeq a$) and spheroidal galaxies correspond to ($ a \simeq b \simeq c$). Conventionally, the shape of galaxies is described by the following shape parameters: the minor-to-major axis ratio, $s \equiv a/c$, as well as the intermediate-to-major axis ratio, $q \equiv b/c$. 
In addition, using the above shape parameters, one can infer 
the triaxiality parameter as $T \equiv (1 - q^2)/(1 - s^2)$, to categorize the galaxy shapes. Here $T = 0 (1)$ refers to a perfect oblate (prolate) spheroid. Conventionally, \cite{2012JCAP...05..030S} defines $T \leq 0.33$ to indicate an oblate ellipsoid, while $T \geq 0.66$ implies a prolate ellipsoid. Finally, triaxial ellipsoids are those in the range $ 0.33 \leq T \leq 0.66$. 
Observational estimates \citep{2006ApJ...652..963R, 2014ApJ...792L...6V, 2019MNRAS.484.5170Z} indicate that galaxies tend to be elongated at low stellar mass and high redshifts and are more disky at higher stellar mass and low redshifts.  

Theoretically, there have been many works trying to estimate the shape of DM halo using different techniques and simulations \citep{1987ApJ...319..575B, 1991ApJ...378..496D, 1992ApJ...399..405W, 1994ApJ...431..617D, 1995ApJ...439..520E, 1998MNRAS.296.1061T, 2019MNRAS.490.4877P, 2002ApJ...574..538J, 2004IAUS..220..421S, 2006MNRAS.367.1781A, 2015MNRAS.453..469T, 2016MNRAS.462..663B, 2020arXiv200503025S}. While these studies confirm the triaxial nature of DM halos, their inferred shape slightly differs. 

Almost all of these studies ignore the impact of baryons in their shape estimation or just include adiabatic hydrodynamics. The general consensus is that the ratio of minor, $a$, to major axes, $c$, for DM halos, without baryonic effects included, is about $s = (a/c) = 0.5 \pm 0.2 $ (see \cite{2004IAUS..220..421S} and references therein). 
Furthermore, we may expect that the DM halo shape in this case depends on the angular momentum of halo mergers.

Including baryonic effects, such as gas cooling \citep{2004ApJ...611L..73K} and star formation, is much more demanding. It is believed that these effects would ``round up'' the galaxies and make a transition from a prolate halo to more oblate/spherical one. \cite{2004ApJ...611L..73K} showed that gas cooling leads to average enhancement of principal axis ratios by an amount of order $\simeq (0.2-0.4)$. 
There have also been several works trying to model the potential and shape of MW halo, in particular, by using stellar streams. For example, \cite{2010MNRAS.407..435A} investigated the radial dependence of halo shapes inside their hydrodynamic simulations and found $s = (a/c) \sim 0.85$. 
On the other hand, \cite{2010ApJ...714..229L} estimated $(a/c) \simeq 0.72$ and $b/c \simeq 0.99$. Another interesting quantity in this context is the alignment of the angular momentum of the galaxy with the minor axis. \cite{2005ApJ...627..647B} found a mean misalignment of order $\simeq 25$ \rm{deg}. \cite{2019MNRAS.484..476C} used the Illustris simulation and showed a median of $s \simeq 0.7$ for halo masses $ M \leq 10^{12.5} M_{\odot}$. \cite{2019MNRAS.490.4877P} used AURIGA simulations to infer DM halo shape and demonstrated that baryonic effects make DM halos rounder at all radii as compared with dark matter only (DMO) simulations \citep{2016MNRAS.458.1559Z}. 

There are also some studies (look for example at \citep{2019MNRAS.488.5580P} and references in there).  trying to use the elongation of the low mass galaxies to shed light on the cosmic web at the high redshifts.

In this paper, we build on previous investigations and use TNG50 \citep{2019MNRAS.490.3196P, 2019MNRAS.490.3234N}, a very high resolution hydrodynamical simulation that simultaneously evolves dark and baryonic matter in a cosmological volume. We estimate the shape of DM halos for a sample of 25 MW-like galaxies. We use three different approaches in calculating the shape parameters. Our main algorithm is based on the standard Enclosed Volume Iterative Method (\rm{EVIM}).  We also infer the shape using a Local Shell Non-Iterative Method (\rm{LSNIM}) as well as a Local Shell Iterative Method (\rm{LSIM}). We compare the radial profile of shape parameters in these approaches at the level of median and percentiles. We classify halos in 3 different categories. Simple, twisted and stretched halos present different features in the radial profile of their axes lengths as well as the level of axes alignment between different radii. We make a comprehensive exploration of different halos and connect the shape with different halo properties as well as the satellites of the halo. To seek for the impact of the baryonic effects, we study DMO simulations, with no baryons. Finally, we study the connection of these theoretical results and some recent observations. 

The paper is organized as follows. Sec. \ref{Method} reviews the simulation setup. Sec. \ref{shape-Analysis} presents different algorithms in computing the DM halo shape. Sec. \ref{shape-analysis} focuses on the shape analysis. Sec. \ref{Driver}
discusses the main drivers of the shape. Sec. \ref{Observation} studies the connection with observations. Sec. \ref{conc} presents the summary of results. Some of the technical details such as non-disky galaxies and convergence check are left to Appendix \ref{non-disk}-\ref{Weighting-factor}.

\section{METHODOLOGY AND DEFINITIONS}
\label{Method}

\subsection{TNG50 Simulation}
TNG50, the third in the series of IllustrisTNG simulations, refers to the highest resolution realization of large-scale hydrodynamical cosmological simulations \citep{2019MNRAS.490.3196P, 2019MNRAS.490.3234N} providing a remarkable combination of volume and resolution as listed in Table \ref{TNG50}. As a hydrodynamical cosmological simulation, it evolves gas, DM, SMBHs, stars and magnetic fields within a periodic-boundary box. The softening length is 0.39 comoving \rm{kpc/h} for $z \geq 1$ and 0.195 proper \rm{kpc/h} for $z <1$. 

Its initial conditions have been chosen from a set of 60 realizations of the initial random density field at $z = 127$ and using the Zeldovich approximation. The choice of its cosmological parameters are based on \cite{2016A&A...594A..13P} with the following parameters, $\Omega_m = \Omega_{dm} + \Omega_b = 0.3089$, $\Omega_b = 0.0486$, $\Omega_{\Lambda} = 0.6911$, $H_0 = 100 h \rm{km} s^{-1} \rm{Mpc}^{-1}$, $h = 0.6774$, $\sigma_8 = 0.8159$ and $n_s = 0.9667$. 
The above initial conditions have been evolved using the \rm{AREPO} code \citep{2010MNRAS.401..791S} which solves a set of coupled equations for the magnetohydrodynamics (\rm{MHD}) and self-gravity. Poisson's equations for gravity are solved using a tree-particle-mesh algorithm.
\begin{table*}[!htbp] 
	\caption{Physical parameters of TNG50 simulation. This includes simulation volume, box side length, number of gas and DM particles, target baryon and DM mass and $z = 0$ Plummer (equivalent) gravitational softening of DM and stars.  }
	\label{TNG50}
	\begin{tabular}{|l|c|c|c|c|c|c|c|c|r|} 
		\hline 
		 \textbf{Name} & \textbf{Volume} [$\left(\rm{Mpc} \right)^3$] & 
        $\textbf{L}_{\rm{box}} [\rm{Mpc}/h]$  &
        $\textbf{N}_{\rm{GAS}}$ &  $\textbf{N}_{\rm{DM}}$ &  $\textbf{m}_{\rm{baryon}}$ [$10^5 M_{\odot}$] & $\textbf{m}_{\rm{DM}}$ [$10^5 M_{\odot}$] & $\mathbf{\epsilon}_{\rm{DM,stars}}$ [\rm{kpc}/h]
        \\
		\hline
    TNG50    &$51.7^3$ & $35$ & $2160^3$ &   $2160^3$ &  $0.85$ & $4.5$ & $0.39 \rightarrow 0.195 $ \\
       	\hline 
       	 TNG50-Dark  &$51.7^3$ & $35$ & $-$ &   $2160^3$ &  $-$ & $5.38$ & $0.39 \rightarrow 0.195$ \\
       	 \hline
       	\end{tabular}
\end{table*}

Models for unresolved astrophysical processes, such as star formation, stellar feedback and SMBH formation, growth and feedback used in TNG50 are the same as the other IllustrisTNG simulations and are described in detail in \cite{2017MNRAS.465.3291W, 2018MNRAS.473.4077P}.

\subsection{Milky-Way like galaxies}
\label{MW-gal-sample}
Halos in TNG simulations are identified using a friends-of-friends (FOF) group finder algorithm \citep{1985ApJ...292..371D}. Furthermore, gravitationally self-bound subhalos are determined based on the \rm{SUBFIND} algorithm \citep{2001MNRAS.328..726S, 2009MNRAS.399..497D}. Every FOF has one central galaxy, the most massive subhalo, and a number of satellites. Here we limit our study to central subhalos with masses in the range $ (1- 1.6) \times 10^{12} M_{\odot}$, which is the halo mass range appropriate to Milky Way like galaxies \citep{2019A&A...621A..56P}. There are 71 galaxies in TNG50 within the above selection criteria. Furthermore, below we add an extra selection criterion and choose disk-like galaxies from the above sample. This is motivated as according to the observations of local Universe, it is manifest that massive star forming galaxies, such as MW galaxy, present a disk-like shape \citep{2013ApJ...779...42S}. 
As described in the following, our selection is based on the orbital circularity parameter. 

\begin{figure*}
\center
\includegraphics[width=0.9\textwidth]{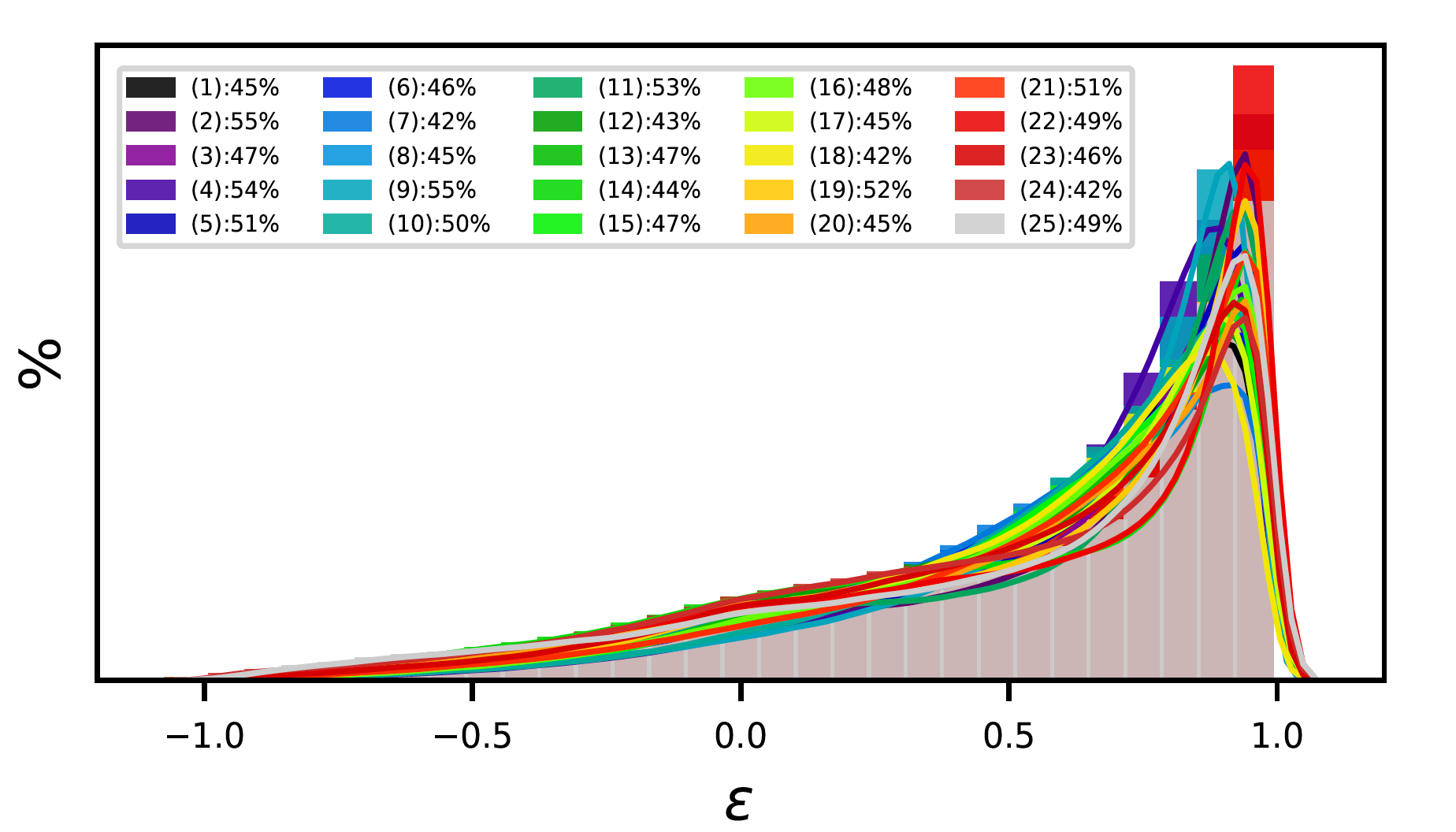}
\caption{Probability density of orbital circularity parameter, $\varepsilon$, for a sample of 25 MW like galaxies in TNG50 simulation labeled according to the disk-frac. Different colors describe various galaxies.  \label{epsilon-TNG50}}
\end{figure*}

Galaxies in the TNG simulation are split into cases with dominant rotationally supported disks and isotropic bulge-like shapes. Throughout this work, we are mostly interested in the former case. In the following, we describe an algorithm to identify rotationally supported stellar components. Our formalism is based on \cite{2003ApJ...591..499A, 2018MNRAS.473.1930E}. 

First, we calculate galaxy's stellar net specific angular momentum vector, $\mathbf{j}_{\rm{net}}$:

\begin{equation}
\label{net-J}
\mathbf{j}_{\rm{net}} \equiv \frac{\mathbf{J}_{\rm{tot}}}{M} = \frac{\sum_{i} m_i \mathbf{r}_i \times \mathbf{v}_i}{\sum_{i} m_i},
\end{equation}
where the index $i$ refers to star particles. Conventionally, the $z$ axis is pointed to the $\mathbf{j}_{\rm{net}}$ direction.

Next, for every star particle, we calculate the component of its angular momentum along the $z$ direction, pointed to the direction of $\mathbf{j}_{\rm{net}}$:

\begin{equation}
\label{net-J1}
j_{z,i} = \mathbf{j}_i \cdot \mathbf{\hat{z}} ~~~,~~~ 
\mathbf{\hat{z}} \equiv \frac{\mathbf{j}_{\rm{net}}}{|\mathbf{j}_{\rm{net}}|} ~~~,~~~ \mathbf{j}_i = \mathbf{r}_i \times \mathbf{v}_i,
\end{equation}
Finally, we define the orbital circularity parameter as the ratio between $j_{z,i}$ and $j_c(E_i)$ (which is the specific angular momentum of $i$-th star particle in a 
circular orbit, with radius $r_c$, and with the same energy as $E_i$):

\begin{equation}
\label{orbital-circularity}
\varepsilon_i \equiv \frac{j_{z,i}}{j_c(E_i)} ~~~,~~~ j_c(E_i) = r_c v_c = \sqrt{G M(\leq r_c) r_c}. 
\end{equation}
For every particle, we compute the radius of the circular orbit by equating the particle's energy with the specific energy of a circular orbit $E(r_c) = \frac{G M(\leq r_c)}{2r_c} + \phi(r_c)$ where $M(\leq r_c)$ refers to the mass interior to the circular orbit. $\phi(r_c)$
refers to the radial profile of an averaged gravitational potential for a collection of stars, gas, DM and central BH in some radial bins within the galaxy. To compute $\phi(r_c)$ we divide the radial distance (from the center) to many different bins and compute the averaged total potential at every location. 

Since the circular orbit has the largest angular momentum, we have $|\varepsilon_i| \leq 1 $ with positive/negative values for prograde/retrograde orbits while $\varepsilon_i =0 $ for radial or isotropic orbits, i.e. bulge-like components. We identify the stellar disk as those star particles with $\varepsilon_i \geq 0.7$, where hereafter we remove indexed $i$ for the brevity. Furthermore, we limit our searches to cases with fraction of stars in the disk (hereafter Disk-frac), defined with $\varepsilon \geq 0.7$, above 40\% located in radial distance less than 10 \rm{kpc} from the center. This ensures us that we have a reasonable fraction of stars in the disk. This criterion reduces the sample to 25 galaxies in TNG50. 
Figure \ref{epsilon-TNG50} presents the distribution of $\varepsilon$ for the sample of MW like galaxies. The dominant prograde orbit of stars is evident from the figure. About 72\% of these galaxies have Disk-frac 
between 40-50 \%, while 28\% 
have Disk-frac between 50-60 \%. 
Since the distribution of orbital circularity parameter has similar profile in our samples, it is intriguing to see what the object-to-object variation in Disk-frac will be for real galaxies. In order to facilitate the presentation of different galaxies, in
Table \ref{halo_ID} we order subhalos and link their ID with a number from 1 to 25. Below we use these numbers instead of halo ID number for brevity.  

\begin{figure*}
\center
\includegraphics[width=0.83\textwidth]{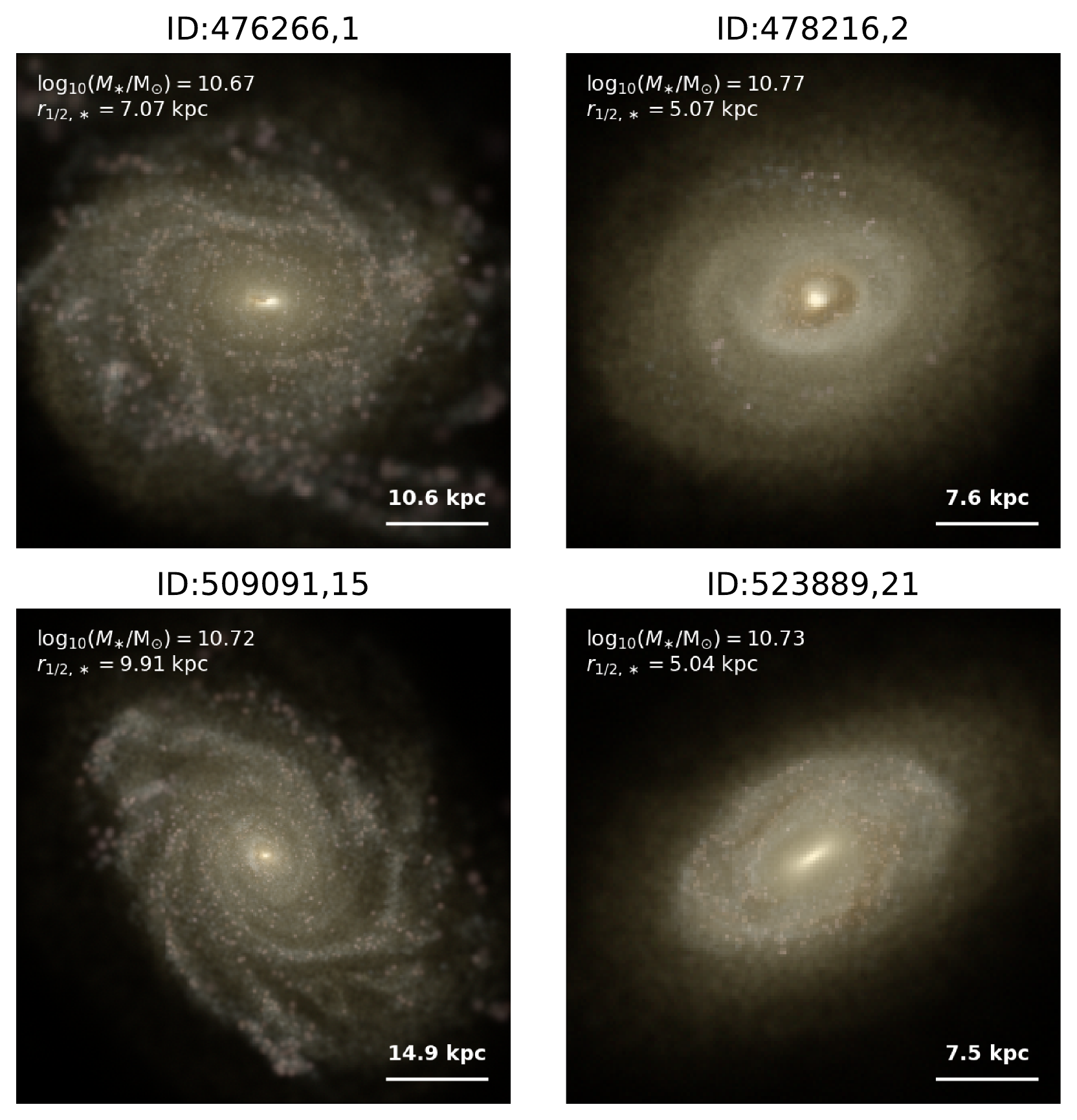}
\caption{Synthetic images of a subset of our sampled MW like galaxies from TNG50 simulation. Images are generated from SKIRT radiative transfer code and using Pan-STARRS1 g,r,i filters. \label{fig:Gal-Image}}
\end{figure*}

\begin{table*}[!htbp] \centering
	\caption{Link between the subhalo ID and the number of halo in MW galaxies.  }
\label{halo_ID} 
\begin{tabular}{|l|c|c|c|c|c|c|c|c|r|} 
\hline 
$1 \mapsto 476266$ & $ 2 \mapsto 478216$ & $ 3 \mapsto 479938$ &   $4 \mapsto 480802$ &  $5 \mapsto 485056 $  \\
\hline 
$6 \mapsto 488530$ & $7 \mapsto 494709 $ & $8 \mapsto 497557$ & 
$9 \mapsto  501208$ & $10 \mapsto 501725 $ \\
\hline 
$ 11 \mapsto 502995 $ & $ 12 \mapsto 503437$ &   $13 \mapsto 505586$ &  $14 \mapsto 506720 $ & $15 \mapsto 509091$ \\
\hline 
$16 \mapsto 510585 $  &
 $17 \mapsto 511303$ & $18 \mapsto  513845$ & $19 \mapsto  519311 $ & $ 20 \mapsto  522983 $ \\
\hline 
 $ 21 \mapsto 523889 $ & $ 22 \mapsto  529365 $ & $ 23 \mapsto  530330$ & $ 24 \mapsto  535410 $ &
 $ 25 \mapsto  538905 $  \\
 \hline 
\end{tabular}
\end{table*}
To get an idea about how different galaxies in our sample look like, in Figure \ref{fig:Gal-Image} we present the synthetic images (using Pan-STARRS1 g,r,i filters) for a subset of our sampled MW like galaxies. The images are taken from the TNG50 Infinite Gallery\footnote{\url{www.tng-project.org/explore/gallery/rodriguezgomez19b/}} at $z = 0.05$ and are matched to our galaxy samples at $z = 0$ using merger trees. 
Each image is generated using the SKIRT radiative transfer code \citep{2015A&C.....9...20C} and includes the impact of dust attenuation and scattering \citep{2019MNRAS.483.4140R}. Text labels indicate the total stellar mass as well as the 3D stellar half-mass radius.

Although from Figure \ref{epsilon-TNG50} the distribution of orbital circularity parameter is fairly similar in all of the galaxies in our sample, their synthetic images show clear morphological differences. It is therefore intriguing to see how does the shape of DM halo differs between these samples.

\begin{figure*}
\center
\includegraphics[width=1.05\textwidth]{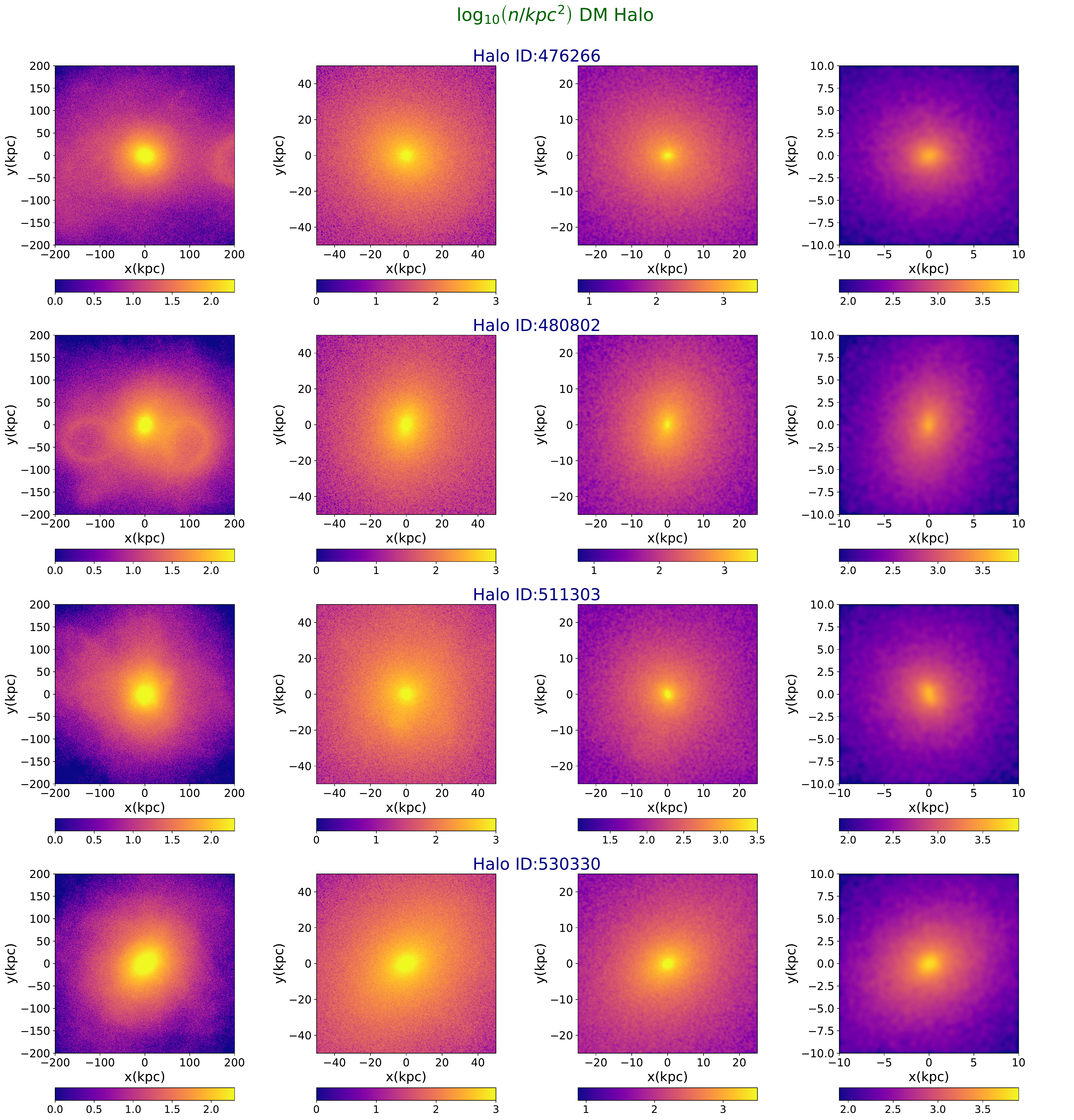}
\caption{Logarithm of the projected number density of DM particles in a sub-sample of MW like galaxies. The halo ID is mentioned on the top of each row. The color bars indicates the projected number density of DM particles. \label{density-xy-dm}}
\end{figure*}

Before we proceed with the analysis of DM halo shapes, in Figure~\ref{density-xy-dm} we analyze the logarithm of 2D projected (number) density of DM particles, in the x-y plane, in a set of 4 MW like galaxies from our sample. Our chosen halos are the same as those in Figure~\ref{fig:Gal-Image}. Each row presents one galaxy, with a given ID number, and from the left to right we zoom-in further down to the central part of the halo. As it is seen from the plot, halo structures varies among different halos. Moreover, in the second row, we see a ghost of a substructure that was imperfectly subtracted from the central halo. 

Below we compute the shape for the above sample (of 25 MW like) galaxies using different techniques. We start with presenting these methods in detail and then infer the shape of dark matter halo accordingly.

\section{Different algorithms in shape analysis}
\label{shape-Analysis}
Having presented the orbital circularity parameter as a key to identifying the 
rotationally supported MW-like galaxies, we turn our attention to DM halo shape. There are different approaches in the literature to computing the galaxy shape \citep[see e.g.][and references therein]{2019MNRAS.484..476C, 2012JCAP...05..030S, 2019MNRAS.485.2589M,2017MNRAS.465.3446G}. Below we compute the DM halo shape using  different methods. While the main focus of our analysis is on a generalization of \citet{2012JCAP...05..030S} method, in Section \ref{local-shell} we adopt the algorithm of \citet{2019MNRAS.485.2589M} and analyze the shape using a non-iterative local shell method. We make a comparison between their final results. To facilitate the reference to these methods, hereafter we name the former method, i.e. enclosed-volume-iterative-method, as \rm{EVIM} and the latter one, local-shell-non-iterative-method, as \rm{LSNIM}. 

\subsection{Enclosed Volume Iterative Method (EVIM)}
\label{EVIM-Shape}

Our main method for the shape analysis relies on the standard algorithm presented in \cite{2012JCAP...05..030S}. Given the triaxial nature of DM halos, we estimate the shape using the axes ratios of a 3D ellipsoid.

We split the interval from $r^i_{\rm{sph}} = 2 $ \rm{kpc} to $r^e_{\rm{sph}} = 202$ \rm{kpc} to $N = 100$ logarithmic radial bins and within each radius, we compute the reduced inertia tensor:
\begin{equation}
\label{shape1}
I_{ij}(\leq r_{\rm{sph}}) \equiv \sum _{n = 1}^{N_{\rm{part}}} \frac{x_{n,i} x_{n,j}}{ R^2_{n}(r_{\rm{sph}})}, ~~~~~ i,j = 1,2,3.
\end{equation}
where $N_{\rm{part}}$ refers to the total number of DM particles interior to an ellipsoid with the axes lengths $(a(r_{\rm{sph}}), b(r_{\rm{sph}}), c(r_{\rm{sph}}))$. Here, $x_{n,i}$ describes the \rm{i}-th coordinate of \rm{n}-th particle. Furthermore, $R_{n}(r_{\rm{sph}})$ denotes the elliptical radius of \rm{n}-th particle, defined in terms of the halo axes lengths as:
\begin{equation}
\label{Elliptical-Radius}
R^2_{n}(r_{\rm{sph}}) \equiv \frac{x^2_n}{a^2(r_{\rm{sph}})} + \frac{y^2_n}{b^2(r_{\rm{sph}})} + \frac{z^2_n}{c^2(r_{\rm{sph}})},
\end{equation}
In this method, our shape computation is based on an iterative approach partially owing to the fact that the elliptical radius contains the axes length ($a(r_{\rm{sph}}), b(r_{\rm{sph}}),c(r_{\rm{sph}})$) which are  unknown {\it a priori}. We shall then determine them in few steps by using an iterative method. 
At any radius, we iteratively compute the reduced inertia tensor starting with all of particles within a sphere of radius $r_{\rm{sph}}$. The eigenvalues and eigenvectors of the diagonalized inertia tensor are then used to deform the initial sphere/(from the second step ellipsoid) while keeping the interior volume fixed. This requires a rescaling of the inferred halo's axes lengths from  $a = \sqrt{\lambda_1}$, 
$b = \sqrt{\lambda_2}$ and $c = \sqrt{\lambda_3}$ to:

\begin{align}
\label{abc}
a = \frac{r_{\rm{sph}}}{(abc)^{1/3}} \sqrt{\lambda_1}, \nonumber\\
b = \frac{r_{\rm{sph}}}{(abc)^{1/3}} \sqrt{\lambda_2}, \nonumber\\
c = \frac{r_{\rm{sph}}}{(abc)^{1/3}} \sqrt{\lambda_3}.
\end{align}

where $\lambda_i, (i = 1,2,3)$ refers to the eigenvalues of the inertia tensor. 
Hereafter we skip showing the radial dependence of $a$, $b$ and $c$ for brevity. Since the eigenvectors of the inertia tensor give us the principal axes, we rotate all of DM particles to the frame of principals as the coordinate system defined with the basis vectors along with the eigenvectors. The only thing to check is that they present a right handed set of coordinates. 
At every step, the halo shape is computed as the ratio of the minor to major axes, $s = a/c$, and the ratio of the intermediate to the major axes, $q = b/c$.

The iteration process is terminated when the residual of both of $s$ and $q$ converges to a level below \rm{max}$(\left((s-s_{\rm{old}})/s\right)^2,\left((q-q_{\rm{old}})/q\right)^2 ) \leq  10^{-3}$ where \rm{max} refers to the maximum between the two quantities. In appendix \ref{Convergence} we present the algorithm in more details. 
After the convergence is established, we also compute the angle between the eigenvectors of the inertia tensor and the total angular momentum of the disk. 

\subsection{Local Shell Non Iterative method (LSNIM)}
\label{local-shell}

Our second approach in computing the halo shape is based on the algorithm of \cite{2019MNRAS.485.2589M}. In this method, the interval between $r^i_{\rm{sph}} = 2.0$ \rm{kpc} to $r^e_{\rm{sph}} = 202$ \rm{kpc}
is divided to 40 spherical shells and we compute the shape at every shell. In addition, we use a non-normalized inertia tensor:
\begin{equation}
\label{shape-tensor2}
\mathcal{M}_{ij} = \sum_{n = 1}^{N_{\rm{shell}}} x_{n,i} x_{n,j},
\end{equation}

where $N_{\rm{shell}}$ presents the total number of particles (of interest) within the shell. Finally, $x_{n,i}$ describes the \rm{i}-th component of the \rm{n}-th particle. Since the particles are considered in spherical shells and as the inertia tensor is not weighted with the elliptical axes, the shape is computed in one step and without any iterations. The above inertia tensor is computed at every spherical shell, with radius $r_i$ and $r_{i+1}$ from the above interval, and is then diagonalized to find the eigenvalues. In addition, since the shells are not deformed, there is not any necessities to rescale the eigenvalues and halo axes lengths are simply the square root of the eigenvalues. 
We shall emphasize here that \rm{LSNIM} is presented only as a comparison and we take \rm{EVIM} as our main approach throughout the subsequent analysis.

\subsection{ Local Shell Iterative method (LSIM)}
\label{local-shell-iterative}
Our third method in analyzing the halo shape relies on a local shell iterative method, \rm{LSIM}. In this approach,  we make 100 algorithmic radial thin shells in the interval between $r^i_{\rm{sph}} = 2 $ \rm{kpc} to $r^e_{\rm{sph}} = 202$ \rm{kpc} and calculate the reduced inertia tensor using Eq. (\ref{shape1}) with the main difference that we replace the interior interval to particles located in thin local shells. At each radius, we iteratively compute $I_{ij}$ in the above shells with an initially spherical shape $a = b = c = r_{\rm{sph}}$ which is distorted iteratively until when the method converges. In a manner similar to EVIM, we compute the eigenvalues and eigenvectors of the localized inertia tensor and deform the shells. Also, to control the deformed ellipsoids locally, at every radius, we take the enclosed volume fixed. The rest of steps are quite similar to EVIM. \cite{2019MNRAS.484..476C} adopted a somewhat similar approach to LSIM in their shape analysis. However, in their analysis, they adopted an unity weighting factor. While for the thin shells we do not expect this change the picture, in Appendix \ref{Weighting-factor} we check this more explicitly and compute the shape parameters using LSIM with two different weighting factors; 1 vs the $r^{-2}$. We find that the final results are relatively insensitive to the choice of the weight. 

 We adopt the EVIM as the main method in our analysis, but we also compare its outcome to that of \rm{LSNIM} and \rm{LSIM} methods in several places of this work. 

\section{Shape profile analysis}
\label{shape-analysis}
Before we proceed with analysing individual halos using \rm{EVIM}, we study the shape profiles at the level of the median/percentiles.

\begin{figure*}
\center
\includegraphics[width=1.0\textwidth]{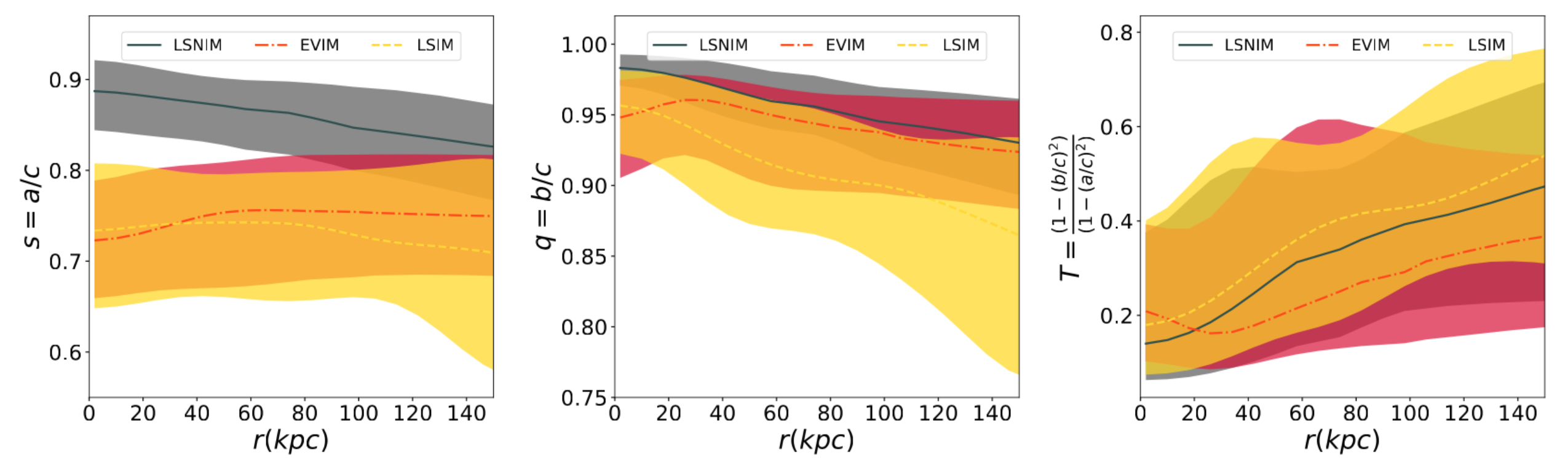}
\caption{ Comparison between the shape parameters $s,q$ and triaxiality parameter T using the \rm{EVIM}, \rm{LSNIM} and \rm{LSIM} algorithms. The sample is stacked over 25 TNG50 galaxies.}  \label{Comparison-s-q-T-DMHalo}
\end{figure*}
\subsection{Statistical shape analysis}
Figure \ref{Comparison-s-q-T-DMHalo} presents the radial profile of the median and 16(84) percentiles of $(s,q,T)$ inferred from 
\rm{EVIM}, \rm{LSNIM} and \rm{LSIM}. From the plot, we can infer several interesting features.

$\bullet$ The radial profile of the $s$ parameter is fairly similar between \rm{EVIM}, \rm{LSIM} in the inner part of the halo but slightly deviates after the radius of 80 \rm{kpc}. On the contrary, \rm{LSNIM} predicts a larger profile for $s$ parameter which is progressively diminishing from the inner to the outer part of the halo.

$\bullet$ 
The inferred radial profile of the $q$ parameter is very close between \rm{EVIM}, \rm{LSIM} up to the radii of 10 \rm{kpc}. However they start deviating from each other after this radius and then \rm{EVIM} gets closer to the \rm{LSNIM} while \rm{LSIM} decreases further out.

$\bullet$ Finally, the inferred radial profile of the 
triaxiality parameter between is fairly close between the \rm{LSIM} and \rm{LSNIM} in terms of the behavior and amplitude but slightly different compared with that of \rm{EVIM}. More explicitly,  while the local method predicts an increasing profile for $T$ throughout the halo, \rm{EVIM} suggests a turn over behavior for $T$ around 30 \rm{kpc.}

$\bullet$ To make the comparison more robust, in Table \ref{Tab:Shape-s-q-T} we present the median and 16th-84th percentiles of halo shape parameters estimated using the above three methods. Here the median and percentiles are computed in two steps. First we compute the radial profile of the median (percentile) using all of galaxies in our sample and then we compute the median (percentile) along the radial direction and up to $r \simeq 200$ \rm{kpc}. The results are rather close to each other. This indicates that the statistical behavior of these approaches are fairly similar.

\begin{table}[h!]	\centering
	\caption{ Median and 16(84)th percentiles of DM halo shape parameters computed from \rm{EVIM}, \rm{LSNIM} and \rm{LSIM}.}
	\label{Tab:Shape-s-q-T}
	\begin{tabular}{lcccr} 
		\hline 
		 Method & 
        $s$ &
        $q$ &  $T$
        \\
		\hline
		\\
       \rm{EVIM} & 
        $0.752_{-0.078}^{+0.062}$
        & $0.944_{-0.048}^{+0.022}$
        & $0.236_{-0.107}^{+0.307}$
        \\
        \\
        \hline
        \\
       \rm{LSNIM} &
       $0.864_{-0.051}^{+0.034}$
        & $0.956_{-0.015}^{+0.022}$
        & $0.346_{-0.198}^{+0.227}$
      \\
      \\
       	\hline
       	\\
       {\color{black} \rm{LSIM}} &
       ${\color{black} 0.726_{-0.073}^{+0.076}}$
        & ${\color{black}0.901_{-0.049}^{+0.041}}$
        & ${\color{black}0.421_{-0.181}^{+0.229}}$
      \\
      \\
       	\hline
	\end{tabular}
\end{table}

Quite interestingly the median of the $s$ parameter is very similar between the \rm{EVIM} and \rm{LSIM} and is smaller than \rm{LSNIM}. On the contrary, the inferred median of $q$ is closer between \rm{EVIM} and \rm{LSNIM} and is slightly larger than that of \rm{LSIM}. Finally, the median of $T$ is minimal from \rm{EVIM} and is maximal from \rm{LSIM}. Although these numbers are consistent with each other at the level of 16(84) percentiles, it is interesting that the median itself shows some levels of sensitivity to the actual method we use.

Finally, the comparison between LSIM and LSNIM is an interesting one.  The fact that both of $(s,q)$ are smaller in LSIM than LSNIM, is intuitively understandable and is related to the fact that non-iterative methods generally predicts more spherical shape profile than the iterative ones.

\subsection{Halo based shape analysis}

Having presented the statistical analysis of the halo shapes, in the following, we analyze the halo shape individually. In addition, hereafter we take the \rm{EVIM} as the main algorithm in our shape analysis. To get a sense of how the results may depend on the actual method, in Appendix \ref{classification}, we compare the shape parameters using both of EVIM and LSIM with each other. Such comparison demonstrates a fair agreement between the results and the halo classifications. Therefore, as already stated above, we use EVIM in our following halo classification.

Based on our shape analysis, we put  DM halos in our galaxy sample in three main categories: (i) Simple, 
(ii) Twisted, and (iii) Stretched halos.
Where halos belong to different classes behave differently in terms of their halo axes lengths as well as the halo orientation. Below we introduce these categories and describe each of them in some depth.

\begin{figure*}
\center
\includegraphics[width=1.0\textwidth]{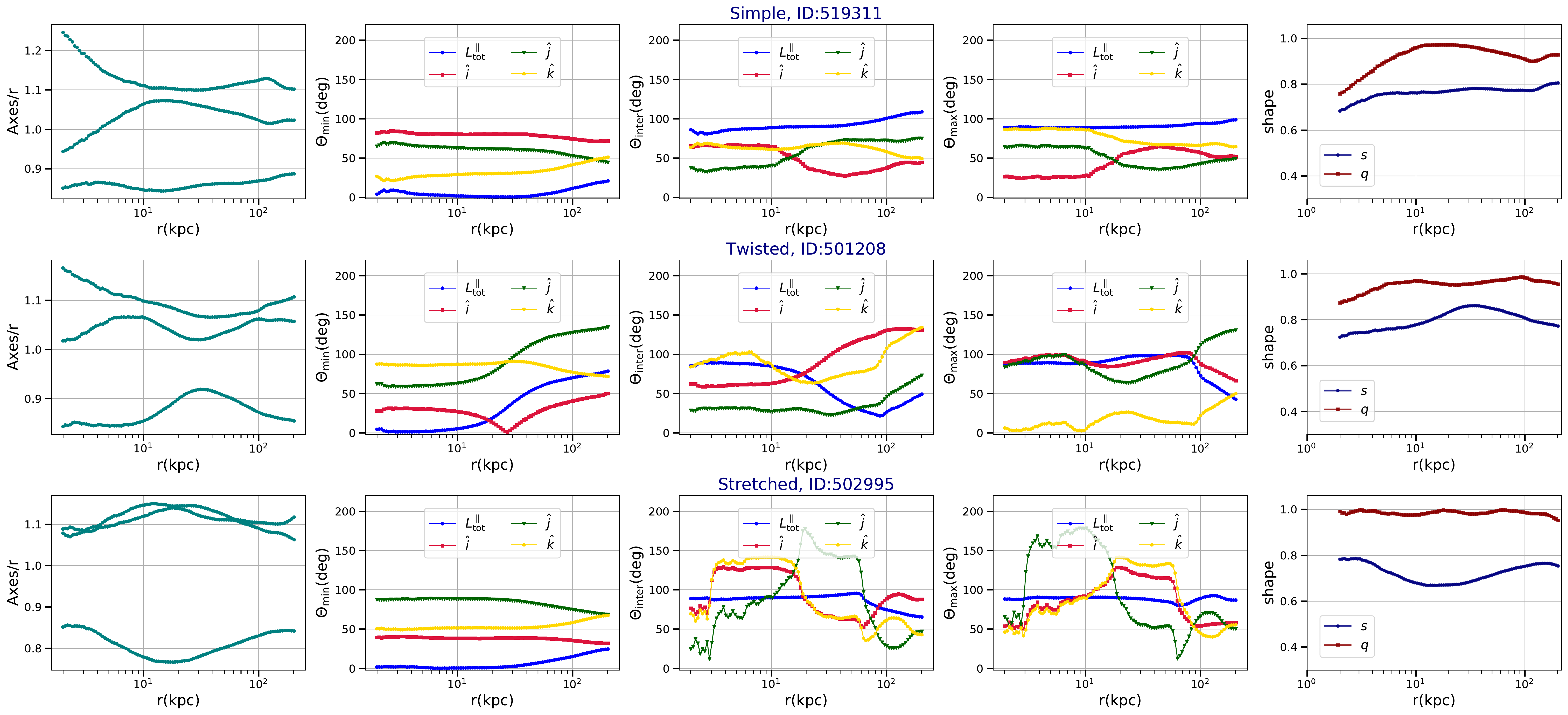}
\caption{The radial profile of 
the axes lengths, angle of \rm{min}, \rm{inter} and \rm{max} eigenvectors with few fixed vectors, $L^{\parallel}_{\rm{tot}}, \hat{i}, \hat{j}$ and $\hat{k}$ and the shape parameters $s,q$ for simple, twisted and stretched halos. Simple halo (first row) presents a small rotation. Twisted halo (second row)  shows some levels of gradual rotation. And stretched halo (third row) establishes a rotation of about 90 degree when the axes lengths cross each other.
\label{simple-twist-stretch}}
\end{figure*}
\subsubsection{Simple Halos}

We start with analyzing the simple halos in our galaxy sample. Based on EVIM, such halos have two main properties; firstly, they have three well separated eigenvalues that the ordering in their magnitude does not change with radius
and secondly, the eigenvector associated with the smallest eigenvalue is almost entirely parallel to the total angular momentum with little change in the angle. In addition, Other eigenvectors present rather small directional variations as well. There are in total 8 halos in this category. In Appendix \ref{classification_1}, we present the radial profile of the axes lengths, as well as the angles of eigenvectors associated with the minimum(\rm{min}), intermediate(\rm{inter}) and maximum(\rm{max}) eigenvectors with different fixed vectors: $L^{\parallel}_{\rm{tot}}$ (the angular momentum of the stellar component of the galaxy) and three unit vectors along the x,y,z directions of the TNG simulation box (refereed to as [$\hat{i}, \hat{j}$, $\hat{k}$]), and finally also the shape parameters $s,q$ for this group.

Although we take the EVIM as our main method, to check the robustness of the properties of the simple halos against using different methods, in Appendix \ref{classification}, we inferred the shape profiles using both of EVIM and LSIM. It is generally correct (with the exception of halo 511303) that the eigenvectors associated with the minimum eigenvalue are almost entirely parallel to the net angular momentum. However, owing to the local fluctuations, in some cases the lines do actually cross each other and a local rotation occurs. Owing to that, we may entitle these halos simple/stretched but just call them simple halos, as for abbreviation.

\subsubsection{\rm{Twisted} halos}
As the second class of halos, here we analyze the twisted halos. Halos belonging to this category show some level of rotation in the radial profile of their eigenvectors. To demonstrate such rotation, we compute the angle of \rm{min}, \rm{inter} and \rm{max} eigenvectors with the aforementioned fixed vectors in 3D such as $L^{\parallel}_{\rm{tot}}, \hat{i}, \hat{j}$ and $\hat{k}$. Twisted halos experience a gradual rotation of about 50-100 \rm{degs} in their radial profile. Furthermore, their axes lengths get very close to each other, sometimes at the corner of crossing, but not really getting stretched, as is described below. There are 8 halos in this class. In appendix \ref{classification_2} we summarize the radial profile of the axes lengths, different angles and shape profile of all of twisted halos.

As it was done in the case of simple halos, in Appendix \ref{classification}, we overlay the shape parameters from the LSIM. Generally speaking, the inferred halo shapes from EVIM and LSIM are fairly close to each other but only slightly noisier in LSIM because of the local fluctuations which may lead to extra crossings of the eigenvalues.

\subsubsection{\rm{Stretched} halos}
As the last class of halos, here we describe the stretched halos. Generally speaking, halos belong to this category experience a level of stretching in the radial profile of their axes lengths. Where two axes lengths approach each other and one of them stretches and gets larger than the other. Thanks to the orthogonality of eigenvectors, such stretching can be easily seen from changing the angle of \rm{min}, \rm{inter} and \rm{max} eigenvectors with the aforementioned unit vectors [$L^{\parallel}_{\rm{tot}}, \hat{i}, \hat{j}$, $\hat{k}$] by about 90 \rm{deg}. Indeed, this criteria allows us to distinguish between the twisted and stretched halos. In summary, the crucial difference between the twisted and stretched halos is that former ones, twisted halos, establish a gradual changes in their angle while the latter one, stretched halos, show more of an abrupt changes in their angles. 

In our galaxy sample, we have 9 halos in this category. In Appendix \ref{classification_3} we summarize the radial profile of the axes lengths, different angles and shape profile of all of stretched halos.

Before we proceed with further study of halo properties in different classes, we shall mention that, there are two halos, halo 4 (ID: 480802) and halo 10 (ID: 501725), which show both of a twisting and stretching in their radial profiles at different radii. More explicitly, these halos demonstrate both of a gradual and abrupt changes throughout their radial profiles and when the principal axis lengths cross each other, respectively. To avoid further complexity of the presentation, however, we put them in the category of stretched halos throughout our following analysis. 

As for the above two cases, in  Appendix \ref{classification_3}, we also overlay the shape profiles from the LSIM. Quite interestingly, the general patterns from EVIM and LSIM are fairly close to each other with extra fluctuations that are arisen owing to the local fluctuations of the profiles.

Having introduced different classes of halos, in Figure \ref{simple-twist-stretch} we present three halos, one from each category, and compare the radial profile of their axes lengths, angles with different vectors and their shape profiles. We use EVIM in our presentation.
It is evident that the level of rotation in the simple halo is much less than the twisted and stretched halos. In addition, while twisted halo, show some level of gradual rotation throughout its radial profile, the stretched halo demonstrates a rotation of about 90 \rm{deg} when the axes lengths cross each other.

\subsection{3D visualization of different halos:}
\begin{figure*}
\center
\includegraphics[width=1.0\textwidth,trim = 6mm 1mm 2mm 1mm]{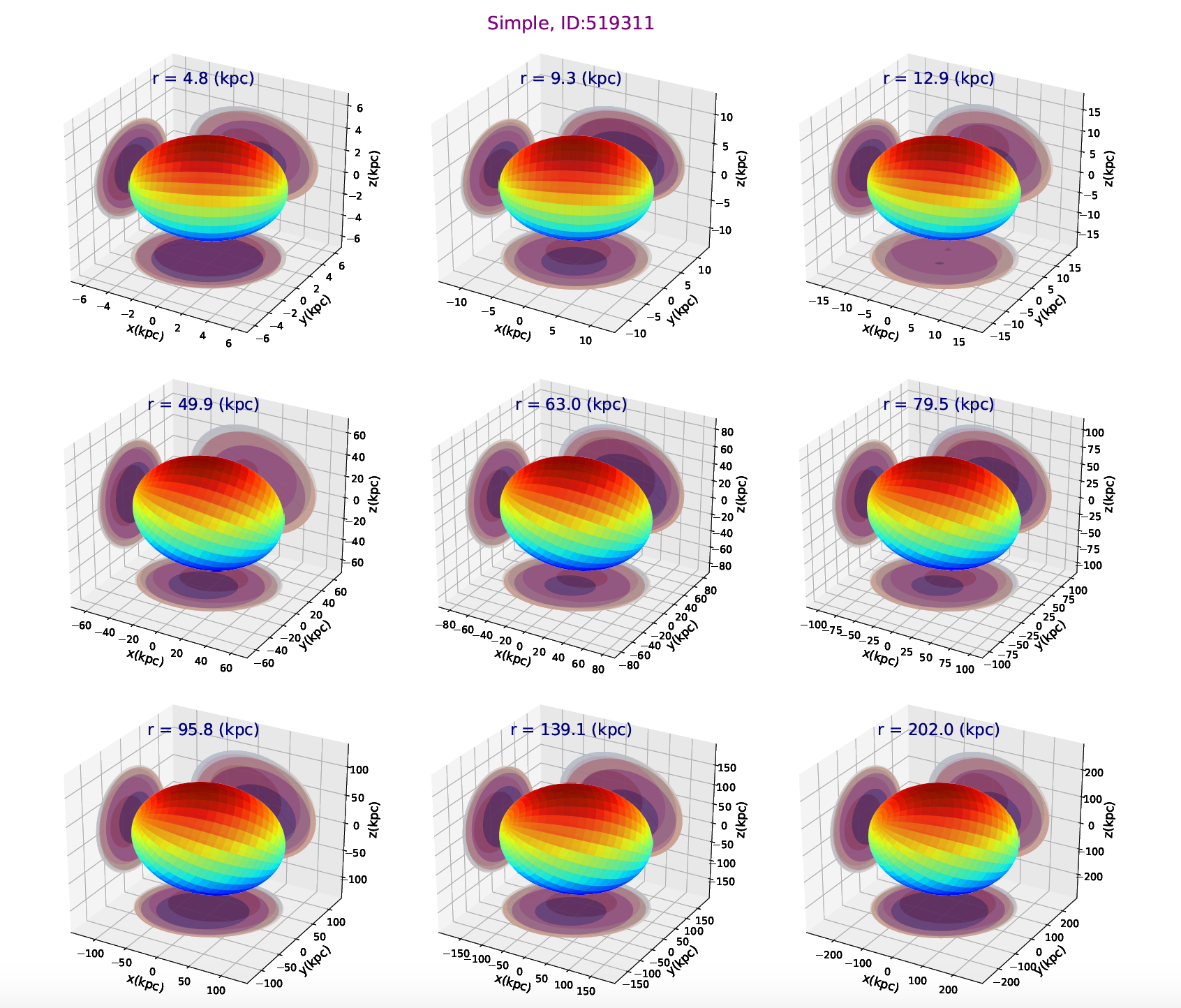}
\caption{3D Ellipsoidal for a Simple halo. The halo experiences a rather weak level of rotation throughout its radial profile.}
\label{Simple-3D}
\end{figure*}

\begin{figure*}
\center
\includegraphics[width=1.0\textwidth,trim = 6mm 1mm 2mm 1mm]{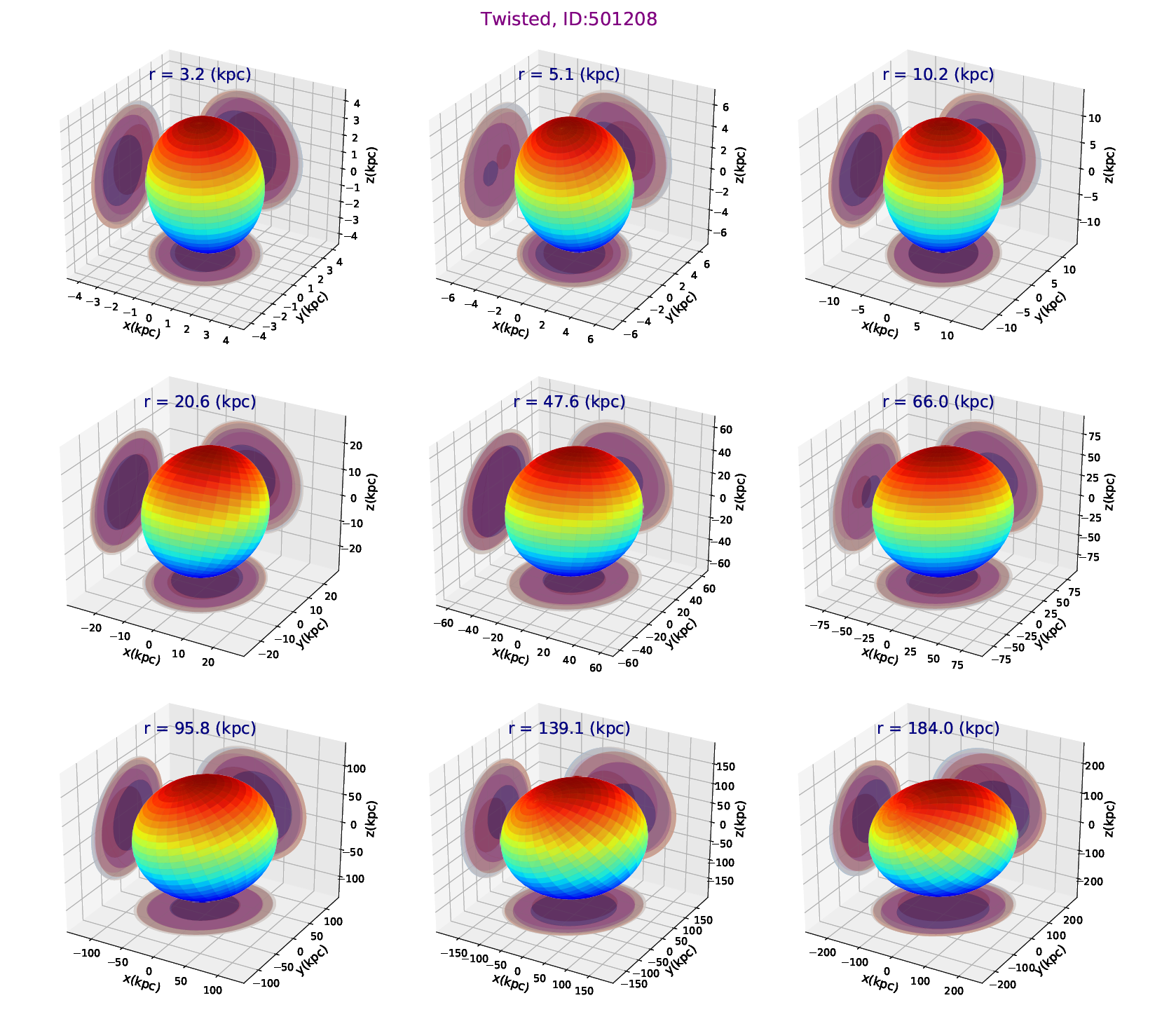}
\caption{3D Ellipsoidal for a twisted halo. The halo establishes some levels of gradual rotations in its radial profile.}
 \label{Twisted-3D}
\end{figure*}

\begin{figure*}
\center
\includegraphics[width=1.0\textwidth,trim = 6mm 1mm 2mm 1mm]{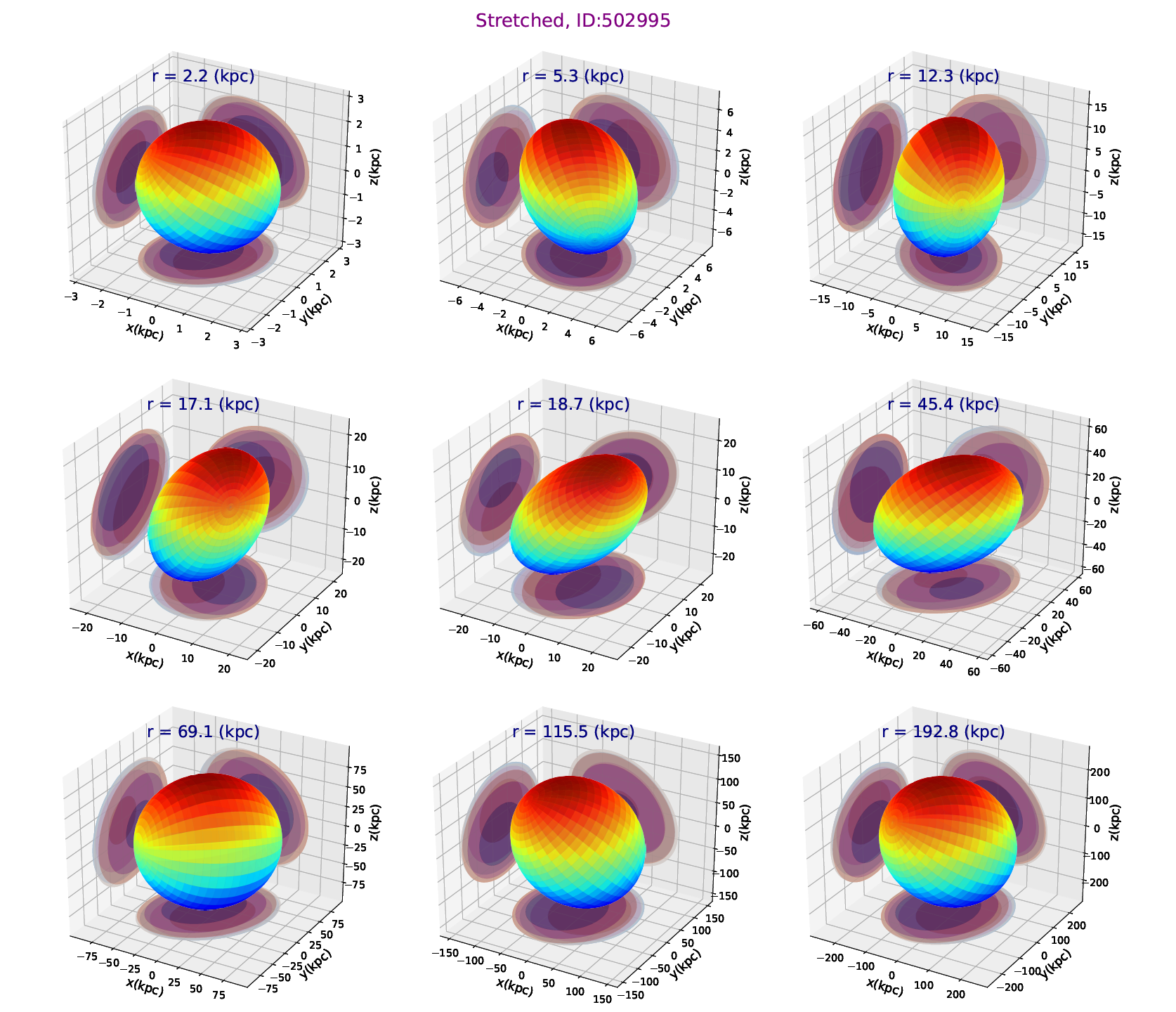}
\caption{3D Ellipsoidal for a Stretched halo. Very close to different crossings, the halo shows large levels of rotations in its radial profile. }
 \label{Stretched-3D}
\end{figure*}

Having specified different halo types in our galaxy sample, here we aim to make a 3D visualization of their ellipsoidal profiles. 
In Figures \ref{Simple-3D}-\ref{Stretched-3D}, we draw 3D ellipsoids of simple, twisted and stretched halos at few different locations. To make each plot, we have ordered the eigenvalues and made a rotation from the principal frame to the Cartesian coordinate using the eigenvectors of the inertia tensor. In addition, at each plot, we also display the 2D projection of 3D ellipsoid in the XY, YZ and XZ planes. 
It is evident that while simple halo presents very little rotation, the twisted and stretched halos establish some level of rotation. In addition, while the twisted halo presents a gradual rotation, the stretched halo establishes a larger rotation near the crossing of the axes lengths.

\subsection{Halo shape and galaxy properties}
\label{halo-property}
Having presented different halo types, below we make the connection between the shape of halos and various galaxy properties such as the galaxy stellar mass as well as the halo formation time.

\subsubsection{Shape vs central galaxy stellar mass}
Here we study the possible connection between the shape parameters and (central) galaxy stellar mass (hereafter galaxy stellar mass), $M_{*}$. Table \ref{Correlation_sqt-M} presents the Spearman correlation between the shape parameters at $r = 2 r_{1/2, *}$ and galaxy stellar mass, where we have chosen a physical distance to draw a better connection with the halo mass. 

Before drawing any conclusions from the table, we should point out that owing to the small sample of simple, twisted and stretched halos, care must be taken in interpreting the statistical significance of these correlations. To take it into consideration, below we not only present the Spearman coefficient, \rm{Coeff}, but we also report the \rm{p-value} for every correlation. 
Furthermore, we name a correlation reliable if its \rm{p-value} is less than 0.05. Having this said, it is clearly seen that only the correlation of $q$($T$) and $M_{*}$ for simple(twisted) halos are meaningful with a positive correlation. The correlation of $T$($q$) and $M_{*}$ for simple(twisted) halos are at the boundary of being reliable.
However, the rest of correlations are not statistically reliable with having larger values of \rm{p-value}.

\begin{figure*}
\center
\includegraphics[width=1.0\textwidth]{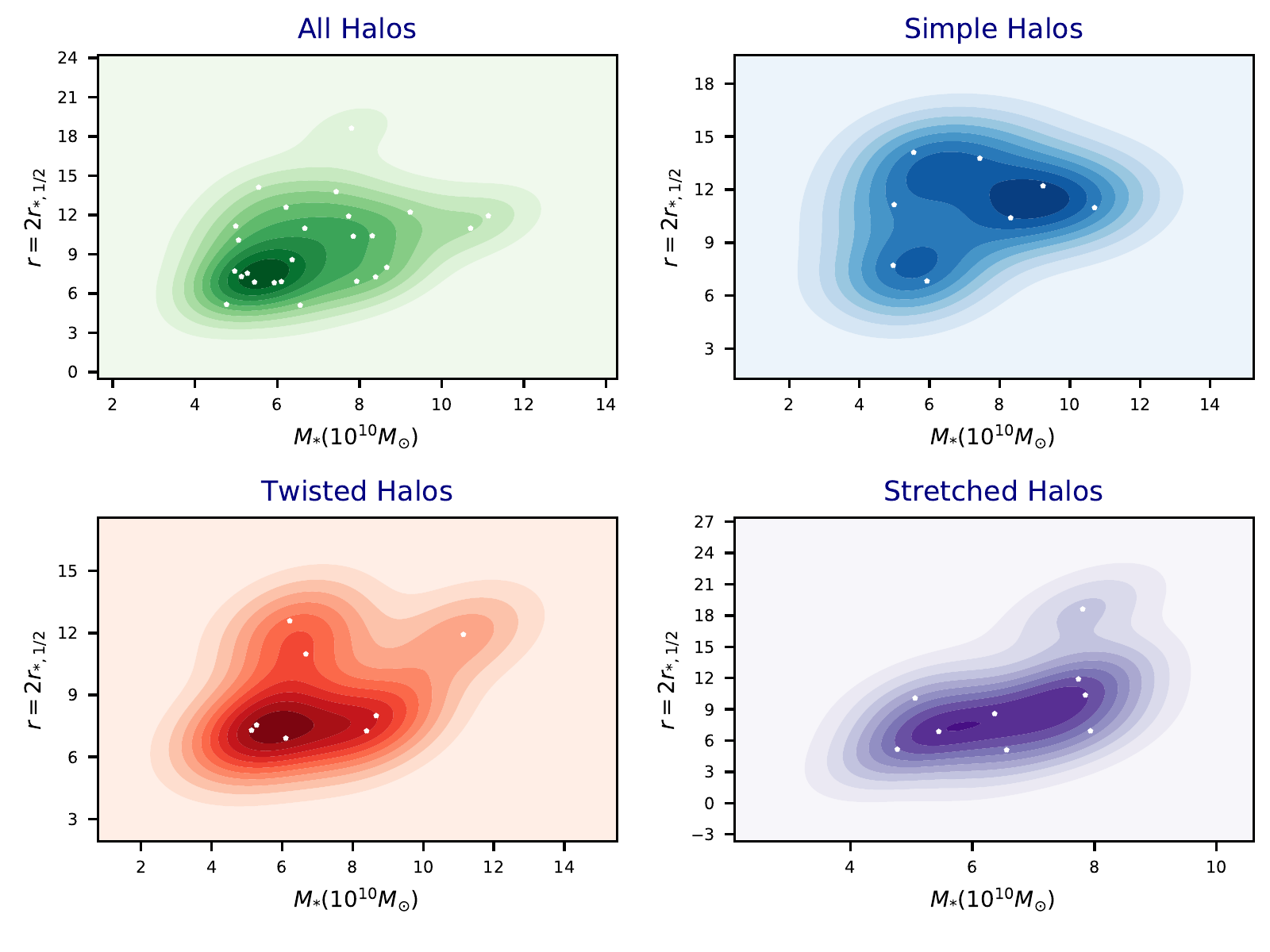}
\caption{Correlation between the halo mass and half mass radius. 
Stretched halos span a narrower range of radii while simple and twisted halos spread over a wider range of masses and radii and they are a bit less compact. Dots in each panel refers to the individual halos.
 \label{mass-radius}}
\end{figure*}

In Figure \ref{mass-radius}, we find the correlation between $2r_{\rm{1/2,*}}$ and $M_{*}$. From the plot, it is evident that stretched halos spans a narrower range of radii while simple and twisted halos spread over a wider range of masses and radii. 

\begin{table}[th!]	\centering
	\caption{Spearman correlation between the shape parameters $(s,q,T)$ located at $r = 2 r_{\rm{1/2,*}}$ and galaxy stellar mass, $M_{*}$.}
	\label{Correlation_sqt-M}
	\begin{tabular}{lcccccr} 
		\hline 
		   & &
        $s$ & $q$ & $T$
        \\
		\hline
       \rm{Simple:} & \rm{Coeff} &
        -0.17
        & 0.90 & -0.71
        \\
       &  \rm{p-value} & 0.69 & 0.002 & 0.05
        \\
        \hline
       \rm{Twisted:} & \rm{Coeff} &
       -0.17
        & 0.57 & -0.64
       \\
      &  \rm{p-value} & 0.69 & 0.14 & 0.09
       \\
       	\hline
       \rm{Stretched:} & \rm{Coeff} &
       -0.017
        & 0.37 & -0.28
       \\
        & \rm{p-value} & 0.97 & 0.33 & 0.46
       \\
       \hline
	\end{tabular}
\end{table}


\subsubsection{Shape vs (stellar) halo formation time}
Below, we make a possible connection between the shape parameters and the halo formation time ($z_{\rm{1/2,*}}$); defined as the time when the half of the galaxy stellar mass is formed. 
Table \ref{Correlation_sqt-z} presents the Spearman correlation between $(s,q,T)$ and $z_{\rm{1/2,*}}$.
As we already mentioned in the correlation of the shape parameters and $M_{*}$, due to the limited sample of halos in each category, we care must be taken in drawing any meaningful conclusions of the real correlation. Owing to this, we also report the \rm{p-value} while presenting the correlation. As it turns out, the only meaningful correlation is between $T$ and $z_{\rm{1/2,*}}$ in the stretched halos. However, we should emphasis here that the sample of stretched halos is very limited with only 5 halos. Being aware of this caveat, we conclude that in our galaxy samples, shape parameters do not seem to be correlated with the halo formation time. 

\begin{figure*}
\center
\includegraphics[width=1.0\textwidth]{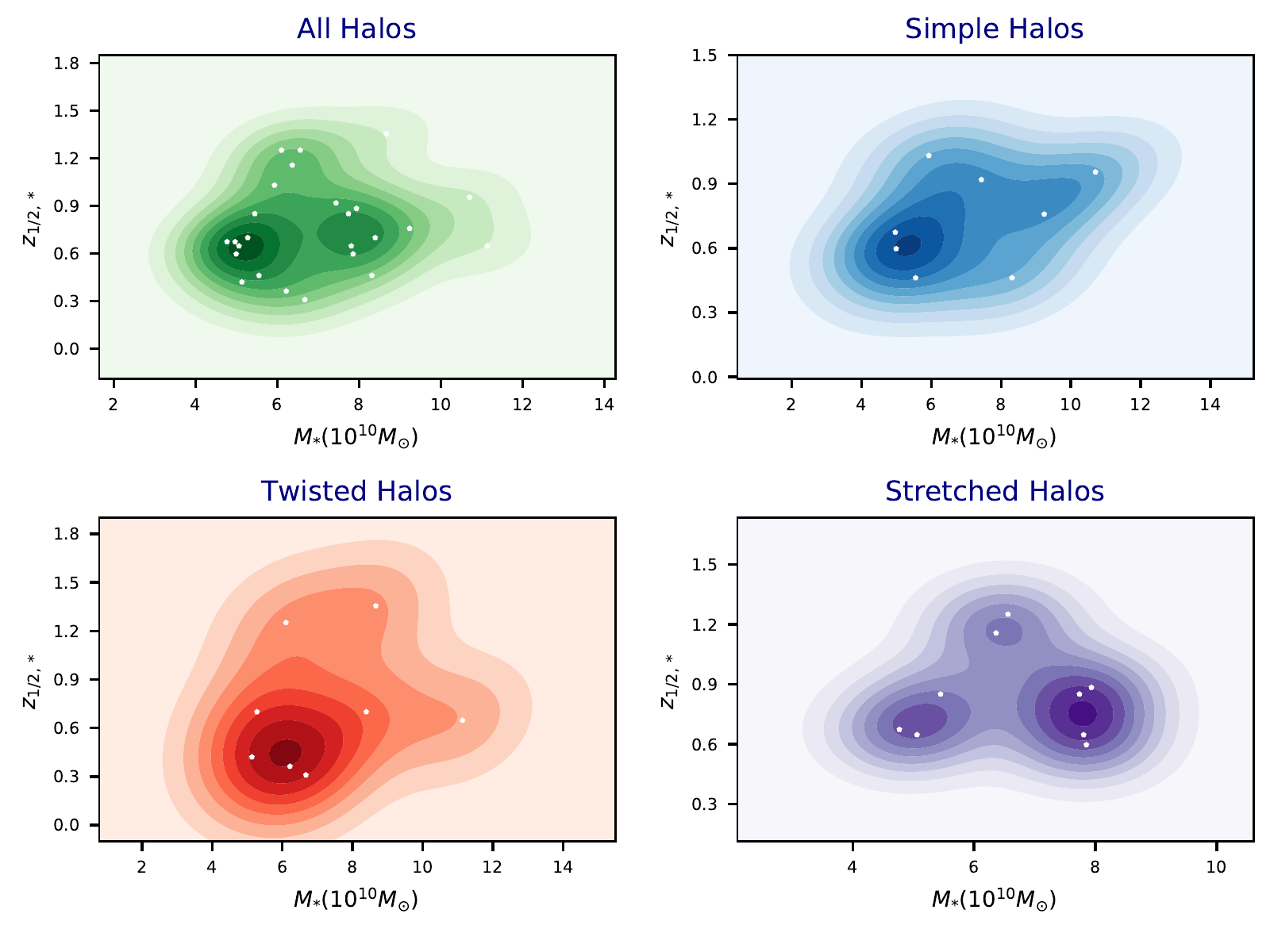}
\caption{Correlation between the halo stellar mass and halo formation redshift. While both of simple and twisted halos show some a positive correlation between $M_{\star}$ and $z_{\rm{1/2,*}}$, stretched halos establish a flat correlation. Dots in each panel refers to the individual halos.
 \label{mass-redshift}}
\end{figure*}

Finally, it is interesting to make a possible connection between the halo mass and its formation time. Figure \ref{mass-redshift} presents the correlation between the galaxy stellar mass as well as the halo formation time. While the simple/twisted halos establish a positive correlation between $M_{\star}$ and $z_{\rm{1/2,*}}$, stretched halos seem to indicate a rather flat correlation. Again we should keep in mind that the sample of stretched halos are a bit limited. Therefore care must be taken in drawing any strong conclusions here.
Being mindful of the statistical caveat here,  
it might be that the nature of different halo types are a bit different. This suggests us to use larger box of TNG simulation and do the analysis for them as well. This is however beyond the scope of this work and is left to a future work. It is also interesting to make a connection between the halo shape and the merger trees. We leave this investigation to a future work as well. 

\begin{table}[th!]	\centering
	\caption{Spearman correlation between the shape parameters $(s,q,T)$ located at $r = 2 r_{\rm{1/2,*}}$ and the stellar halo formation time, $z_{\rm{1/2,*}}$.}
	\label{Correlation_sqt-z}
	\begin{tabular}{lcccccr} 
		\hline 
		   & &
        $s$ & $q$ & $T$
        \\
		\hline
       \rm{Simple:} & \rm{Coeff} &
        0.32
        & 0.22 & 0.12
        \\
        &  \rm{p-value} & 0.43 & 0.61 & 0.78
        \\
        \hline
       \rm{Twisted:} & \rm{Coeff} &
        0.048
        & -0.44 & 0.24
       \\
       &  \rm{p-value} & 0.91 &  0.27 &  0.57
       \\
       	\hline
       	\rm{Stretched:} & \rm{Coeff} &
         0.45
        &  -0.62 &  0.70
       \\
       &  \rm{p-value} &  0.23 & 0.07 &  0.04
       \\
       	\hline
	\end{tabular}
\end{table}


\begin{figure*}
\center
\includegraphics[width=1.05\textwidth]{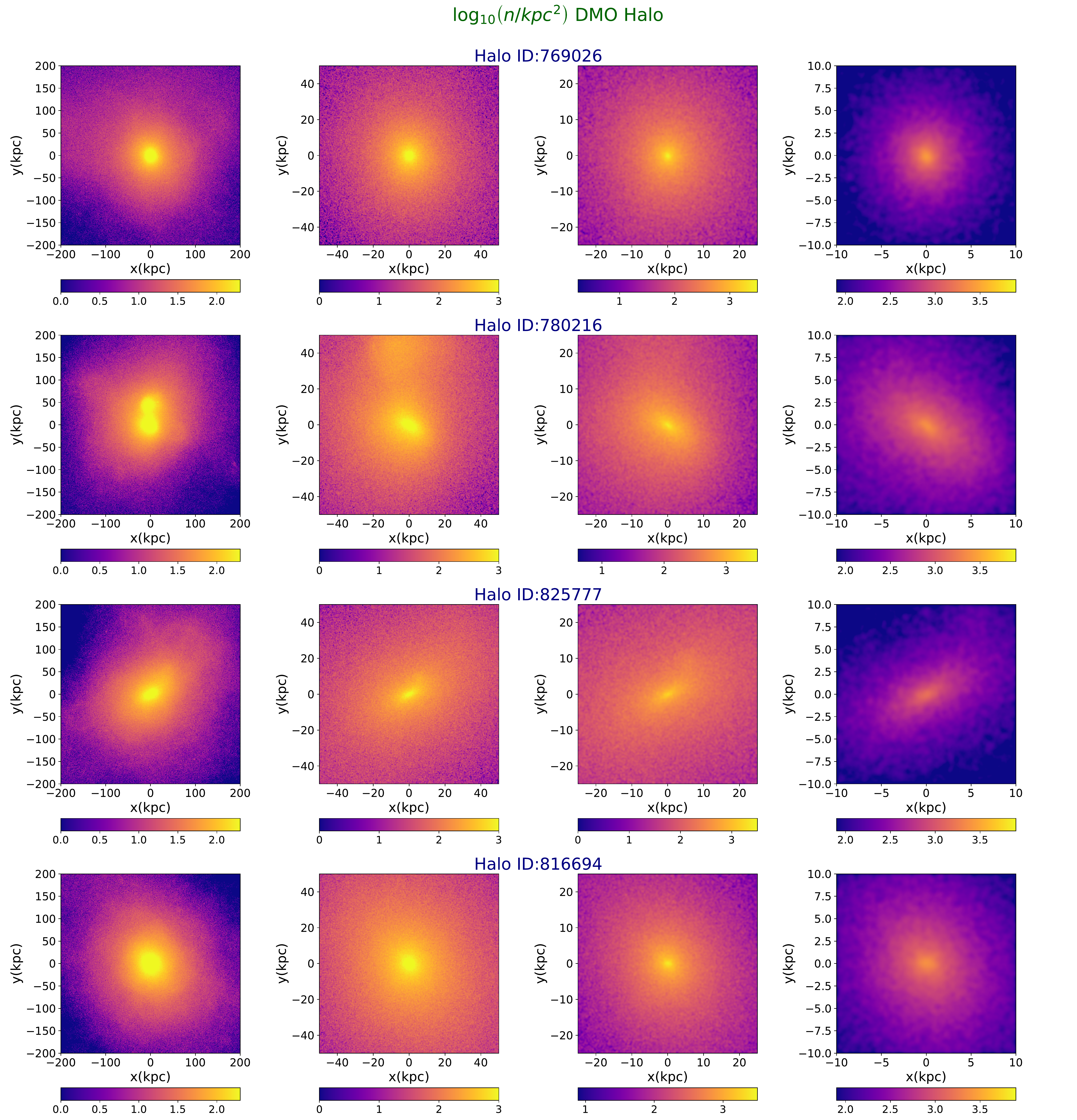}
\caption{Logarithm of the projected number density of DM particles in DMO run of MW like galaxies. The halo ID is mentioned on the top of each row. \label{density-xy-dmo}}
\end{figure*}

\section{Main drivers of the shape}
\label{Driver}
Having computed the shape of the DM halo in MW-like galaxies, below we study the impact of a few different drivers of the shape. Here we mainly focus on three different drivers including baryonic effects and the impact of substructures on halo morphology. In addition, we find the connection between the angular momentum of satellites and the eigenvectors of the reduced inertia tensor as well as $\mathbf{j}_{\rm{net}}$. 

\subsection{Baryonic Effects: DMO simulations}
One possible driver of DM halo shape is baryonic effects. Since baryonic contributions are generally very complex and hard to account for individually, to have an appropriate consideration of their possible impacts, we compute the shape of Dark Matter Only (DMO) simulations similar to \cite{1991ApJ...378..496D,1992ApJ...399..405W,2011MNRAS.415L..69J, 2012JCAP...05..030S, 2019MNRAS.484..476C}. We then compare the final results with the estimated shape from the full hydrodynamical simulations.  

We use TNG50-Dark and compute the shape of DMO MW-like halos. To match the halos from TNG50 to TNG50-Dark, we use the unique ID of dark matter particles. For every galaxy in our preselected sample, we scan the halos in DMO and look for a halo with highest fraction of matched DM particles. It is worth mentioning that TNG50 halos are less massive because of baryonic feedback processes, which eject some mass outside of the halo. This does not occur in the DMO simulation and thus halos in DMO are slightly more massive. To take this into account, in our matching, we take the mass range in TNG50-Dark to be slightly above the original selection in TNG50 as we also have put a prior choice on the range of halo mass to narrow down the interesting mass range a little bit. Therefore the matched DMO halos are found to be more massive; see Table \ref{TNG50} for more details. In most cases, the fraction of matched particles is more than 70\%. In one case, though, \rm{4}th galaxy in the above sample, the percentage of matched DM particles was about 54\%. 

Before proceeding with the shape analysis, in Figure \ref{density-xy-dmo} we study the logarithm of the 2D projected (number) density of DM particles, in (x-y) plane, in a sample of 4 DMO halos of MW-like galaxies under consideration. Our chosen halos are associated with the matched cases to those in Figure~\ref{density-xy-dm}. Each row presents one galaxy, with its given ID number, and from the left to right, we zoom-in further down to the central part of the halo. As expected, the halo structure varies for DMO case as compared with the DM profiles presented in Figure~\ref{density-xy-dm}. 
Although all of the halos in our sample share the same gravity and subgrid baryonic physics, they differ in their formation histories. Owing to this, their density profile differs from each other. For example, the halo in the second row in Figure~\ref{density-xy-dm} is currently experiencing a halo merger. 

\begin{figure*}
\center
\includegraphics[width=1.0\textwidth, trim = 6mm 2mm 0mm 2mm]{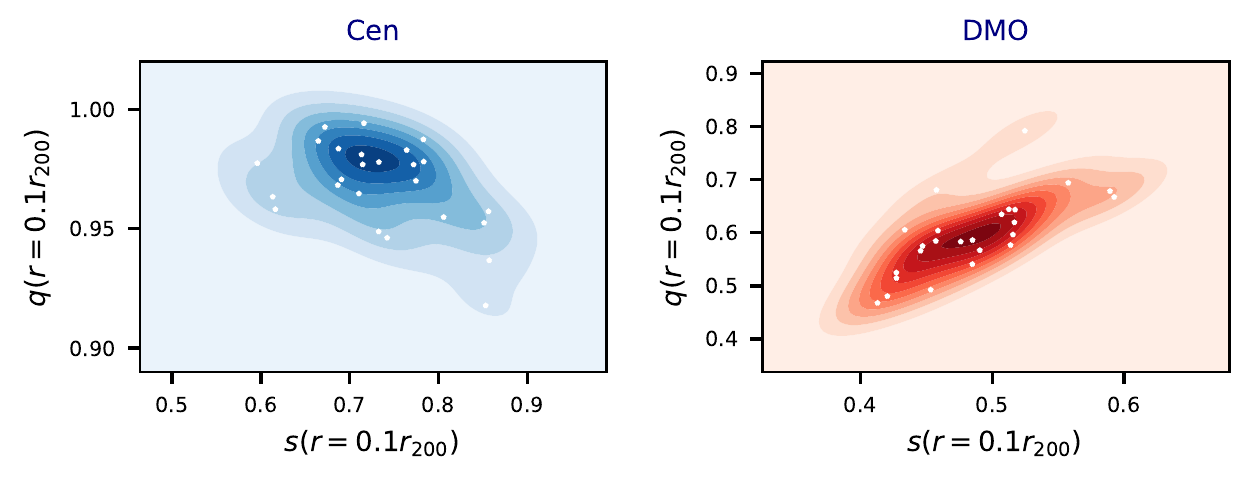}
\caption{Correlation between $s$ and $q$ for different MW like galaxies at $r = 0.1 \rm{R_{200}}$ for the central DM(left) and DMO(right) simulations. Overlaid on the plots, are the scatter plot of $s$ vs $q$.
\label{s-q-r01R200} }
\end{figure*}

In Figure \ref{s-q-r01R200} we draw 
the correlation between $s$ and $q$ ( as inferred from EVIM) at $r = 0.1 R_{\rm{200}}$ for both of TNG50 (Cen:left) versus TNG50-Dark (DMO:right).
Where to distinguish between the central halo and FoF group that would be added later, we explicitly mention Cen here. Overlaid on the plots, are the scatter plot of the same quantities in both cases. It is evident that on average, central halos have larger values of $s$ and $q$ than DMO ones.

\begin{figure*}
\center
\includegraphics[width=1.0\textwidth]{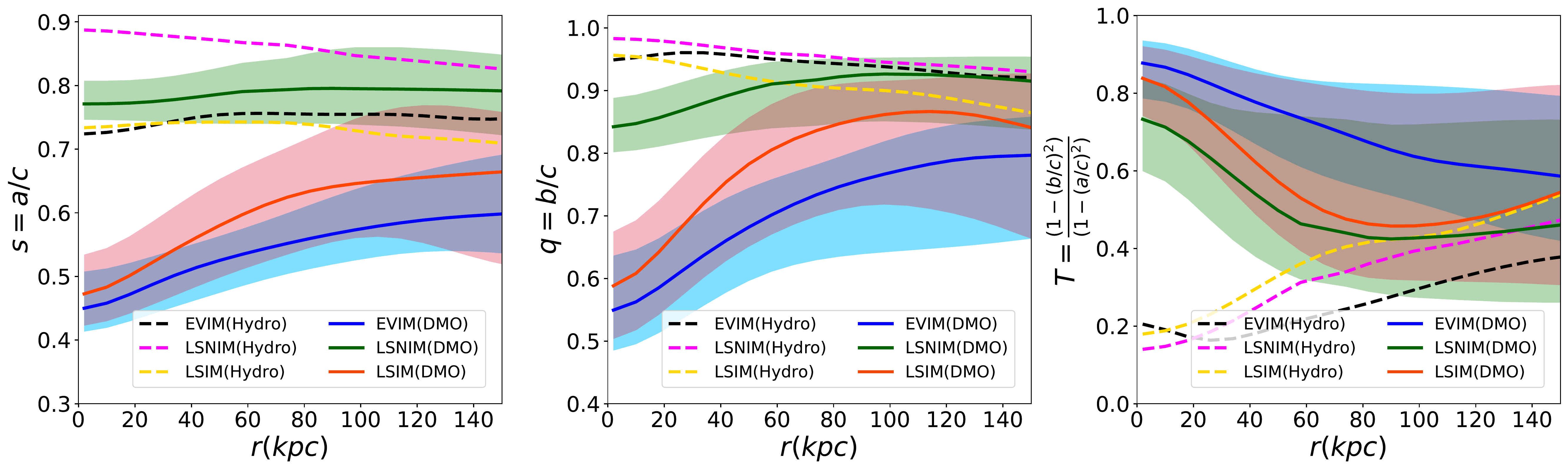}
\caption{Comparison between the shape parameters $s$, $q$ and $T$ for DM halo in DMO computed from \rm{EVIM}, \rm{LSNIM} and \rm{LSIM}. Overlaid on the plot is the median of the shape parameters from the hydrodynamical simulation (Hydro) as well.
\label{Comparison-s-q-T-DMO}}
\end{figure*}

Figure \ref{Comparison-s-q-T-DMO} presents the radial profiles of the median and 16 (84) percentiles of $(s,q,T)$ for DMO simulations inferred using \rm{EVIM}, \rm{LSNIM} and \rm{LSIM}. In contrast to the full-hydrodynamical simulation,  here the inferred $s$ and $q$ from \rm{EVIM}, \rm{LSIM} have radially progressively increasing profiles. Consequently, the radial profile of the triaxiality parameter decreases in both approaches though its actual value is larger in \rm{DMO} compared with the full-hydrodynamical simulations. In Table \ref{Shape-s-q-T-DMO}, we present the median and 16 (84) percentiles of the shape and triaxiality parameters. Comparing this with Table \ref{Tab:Shape-s-q-T}, it is evident that $s$ and $q$ are reduced while $T$ is increased (substantially) in DMO simulation (in \rm{EVIM}). 

\begin{table}[th!]	\centering
	\caption{Median and 16(84) percentiles of DMO shape parameters computed from \rm{EVIM}, \rm{LSNIM} and \rm{LSIM}.}
	\label{Shape-s-q-T-DMO}
	\begin{tabular}{lcccr} 
		\hline 
		 Method & 
        $s$ &
        $q$ &  $T$
        \\
		\hline
		\\
       \rm{EVIM} & 
        $0.571_{-0.053}^{+0.063}$
        & $0.764_{-0.123}^{+0.052}$
        & $0.635_{-0.103}^{+0.188}$
        \\
        \\
        \hline
        \\
       \rm{LSNIM} &
       $0.789_{-0.049}^{+0.042}$
        & $0.912_{-0.075}^{+0.034}$
        & $0.488_{-0.199}^{+0.249}$
      \\
      \\
       	\hline
      \rm{LSIM} &
       $ 0.645_{ -0.130}^{ +0.097}$
        & $0.825_{- 0.164}^{ +0.091}$
        & $ 0.538_{-0.221}^{+0.282}$
      \\
      \\
       	\hline	
	\end{tabular}
\end{table}

Our results suggest that baryonic effects make DM halos more oblate. This result are in great agreement with the conclusion of \cite{2019MNRAS.484..476C}. On the contrary, DM halo looks more prolate for DMO in all of these approaches at small radii and gets more triaxial/oblate at larger radii. Furthermore, $T$ is the largest in \rm{EVIM} and smallest in \rm{LSNIM}.

In conclusion, baryonic effects play a crucial contribution in shaping the DM halo in MW like galaxies. 

\begin{figure*}
\center
\includegraphics[width=1.0\textwidth]{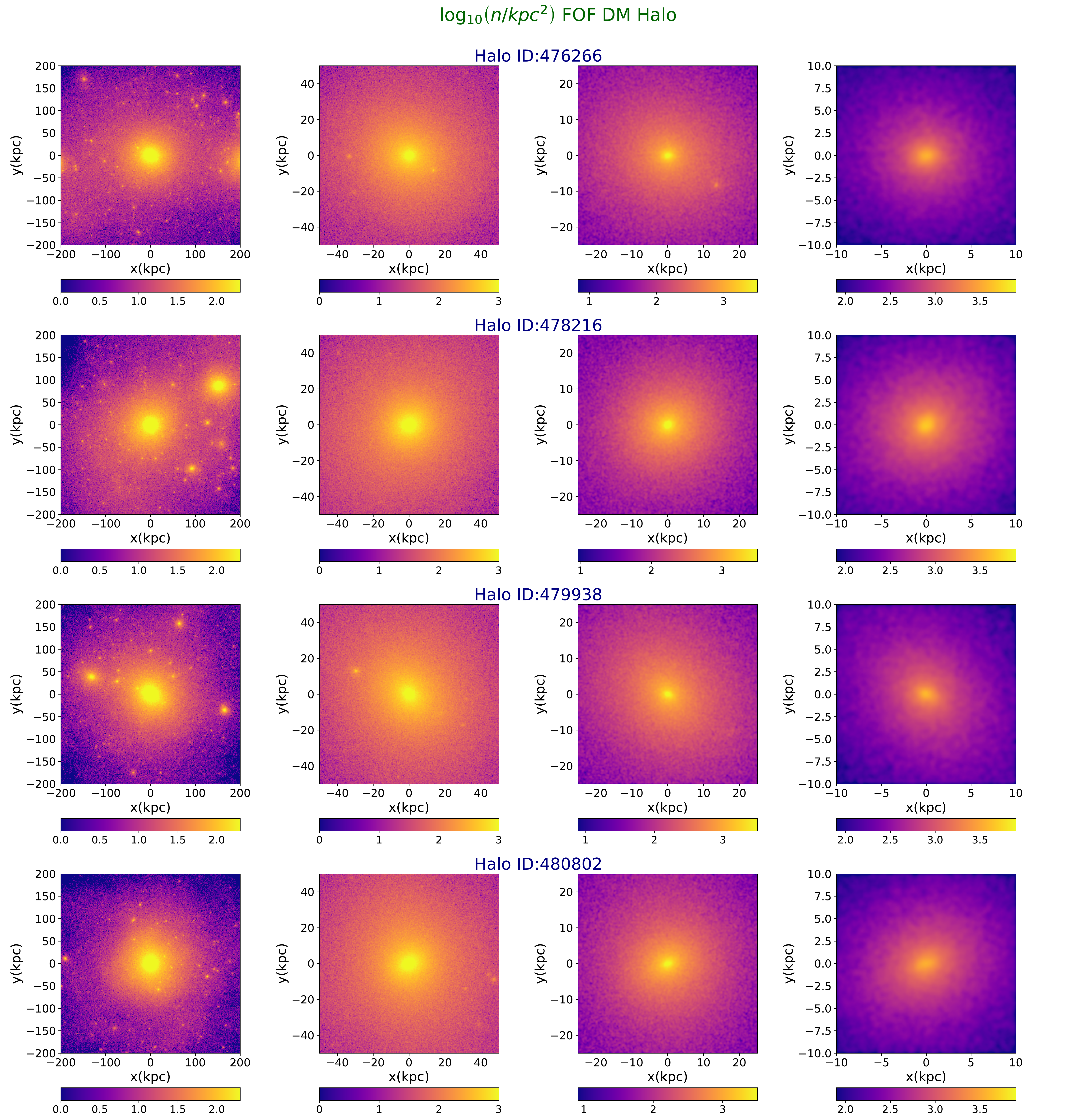}
\caption{Logarithm of the projected number density of DM particles in FoF group. The halo ID is reported on the top of each row. \label{x-y-density-FoF}}
\end{figure*}

\subsection{Impact of FoF group on galaxy morphology}
\label{fof-group}
So far we only studied the impact of central halos in our analysis. Below we generalize our consideration and analyze the impact of different substructures, by using all particles associated with FoF groups in the shape of DM halo. This means that we analyze all particles, including the particles belonging to the central galaxy as well as substructures in our computations. 

First, we study the impact of including all FoF particles on the projected density diagram. In Figure \ref{x-y-density-FoF} we display the logarithm of the projected x-y (number) density profile for FoF group and a sub-sample of 4 galaxies of interest. As expected there are many substructures in the FoF. It is therefore intriguing how they could potentially affect the shape of the DM halo.

\begin{figure*}
\center
\includegraphics[width=1.0\textwidth]
{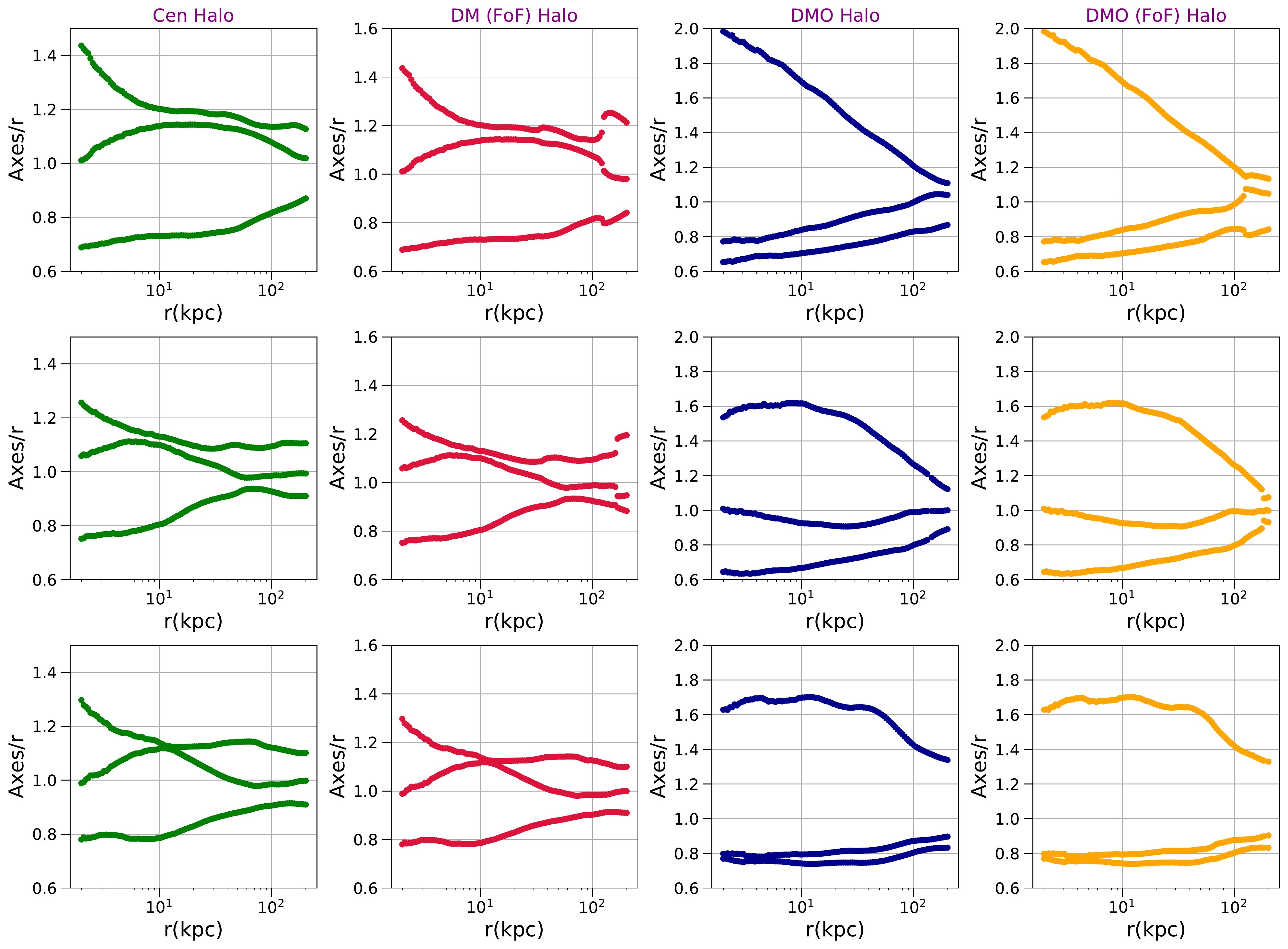}
\caption{ The ratio of Axes lengths to the radius (r) for a sub-sample of MW like galaxies in TNG50. From the left to right, we present the Axes/r ratio for central DM (Cen), DM (FoF), DMO and DMO(FoF)runs of TNG50. While the calculation including all FoF particles is fairly close to the case of the central halo only, DMO simulations look substantially different.}
\label{axes}
\end{figure*}

\begin{figure*}
\center
\includegraphics[width=1.0\textwidth,trim = 6mm 1mm 0mm 1mm]
{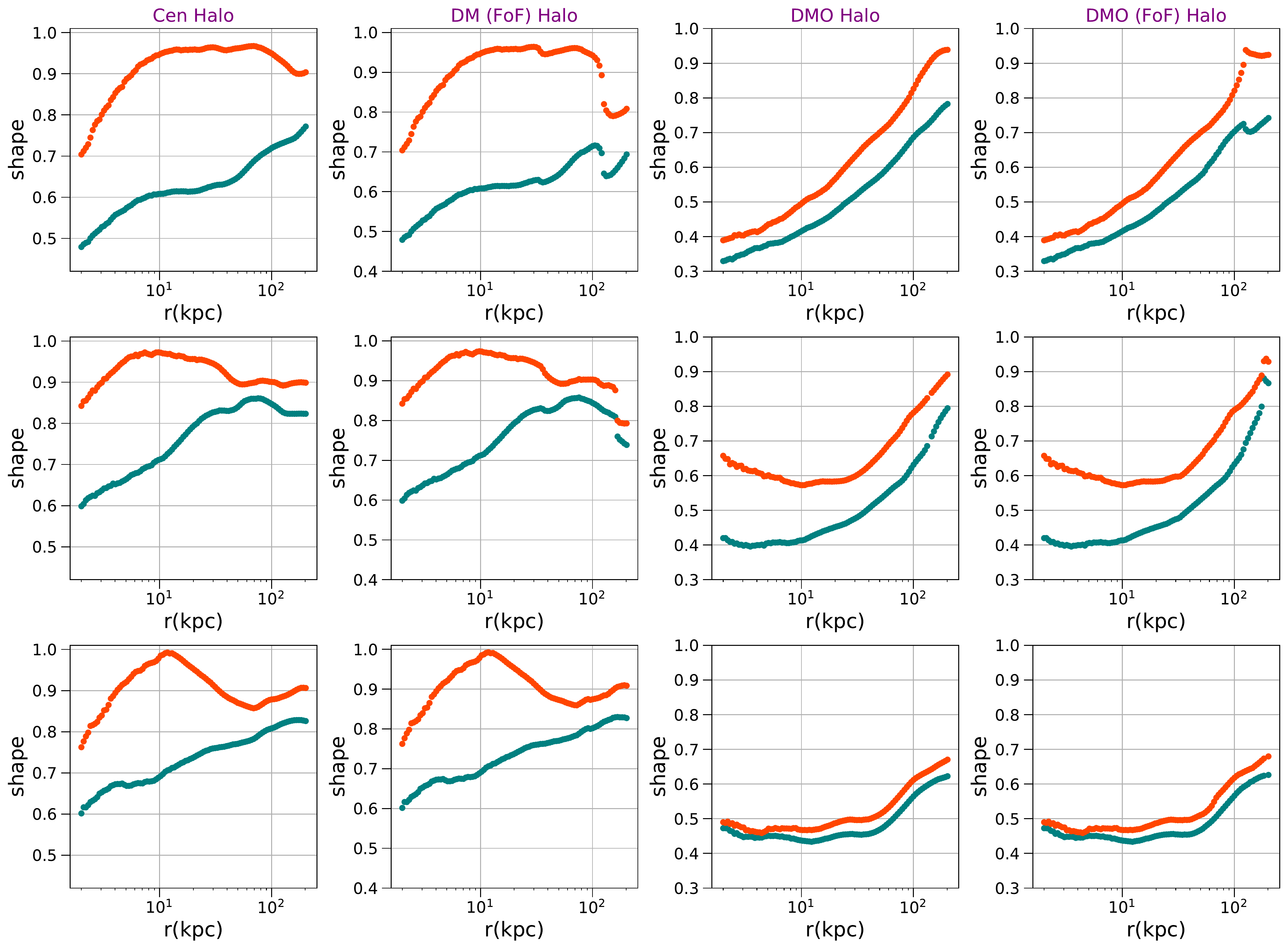}
\caption{ Radial profile of shape parameters for a sub-sample of MW like galaxies in TNG50. From the left to right, we present the radial profile of the shape in DM (Cen), FoF, DMO and DMO (FoF) simulations.}
\label{halo-s-q-dmo-fof}
\end{figure*}

Having presented the central, DMO and FOF group, it is intriguing to compare their shape profiles. To facilitate the comparison, in Figures \ref{axes} and \ref{halo-s-q-dmo-fof}, we present the Axes/r ratio and the shape parameters in all of the above three cases. To make the plots easier to read, we only show the results from our main algorithm, \rm{EVIM}.

Figure \ref{axes} presents the Axes/r ratio for a case of 3 MW-like galaxies in our galaxy sample. In each row, from left to right, we present Axes/r ratio for DM, FoF group, DMO and DMO (FoF) simulations. There are few take aways that can be inferred from the figure. First of all, it is evident that, in inner part of the halo, the impact of substructures is subdominant for both of the hydro and DMO simulations. Next, the radial profile of Axes/r ratio are largely different between the full hydro and DMO simulations. This brings us to the picture that baryonic effects are more important in shaping the halos than the current substructures.

Figure \ref{halo-s-q-dmo-fof} shows the radial profile of the shape parameters $s,q$ for a sample of 3 MW like galaxies from our sample. Overlaid on every plot, we present the shape for DM, FoF group, DMO and DMO (FoF) simulations. The shape profile of FoF group is fairly close to the case of DM halo in the inner part of halo, up to 100 \rm{kpc}. On the contrary, DMO simulation predicts a somewhat smaller value for the shape parameters.

\subsection{Connection with Satellites}
Having presented the Milky Way like galaxies as central subhalos in every group, here we discuss the satellites in individual galaxies. Satellite galaxies are associated with the subhalos which are themselves the members of their parent FoF halo. In IllustrisTNG simulations, subhalos are ordered in terms of their masses, with heaviest member at the beginning of every group identified as the central galaxy; while the rest are assigned as satellites. 

We adopt a stellar mass cut, hereafter $M_{\star}$, in the mass range $M_{\star}\geq 10^7 M_{\odot}$ to ensure that we study subhalos with more than  $\simeq 100$ stellar particles taking into account that the unit mass of baryons in TNG50 is $0.85 \times 10^5 M_{\odot}$. In addition, we restrict ourselves to distances less than 200 \rm{kpc}.
Furthermore, we also eliminate satellites that do not have a cosmological origin. In the TNG simulation, it is done by checking the ``SubhaloFlag".

\begin{figure}
\center
\includegraphics[width=0.45\textwidth,trim = 6mm 2mm 0mm 1mm]{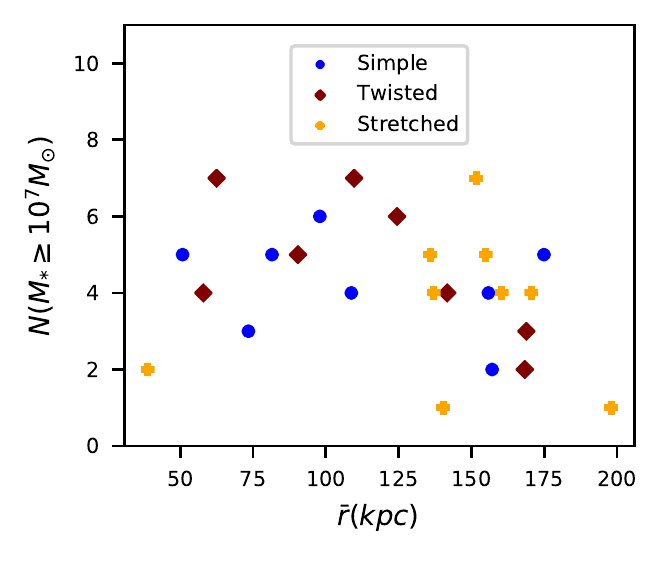}
\caption{Number of Satellites with stellar mass greater than $10^7 M_{\odot}$ (and at the distance less than 200 \rm{kpc}) vs the median distance of the satellites in this mass range from the center.
\label{Nsatellites} }
\end{figure}

In Figure \ref{Nsatellites}, we present the number of satellites within the above mass and distance ranges as the function of their median distance from the halo center. While more than the half of the satellites of simple and twisted halos are in average  closer than 125 \rm{kpc}, those associated with the stretched halo are mostly farther out. 
Furthermore, stretched halos have slightly less satellites compared with the simple and twisted halos. Therefore, care must be taken when we draw a statistical conclusion about the current sample. 

\begin{figure*}
\center
\includegraphics[width=1.0\textwidth,trim = 5mm 2mm 2mm 1mm]
{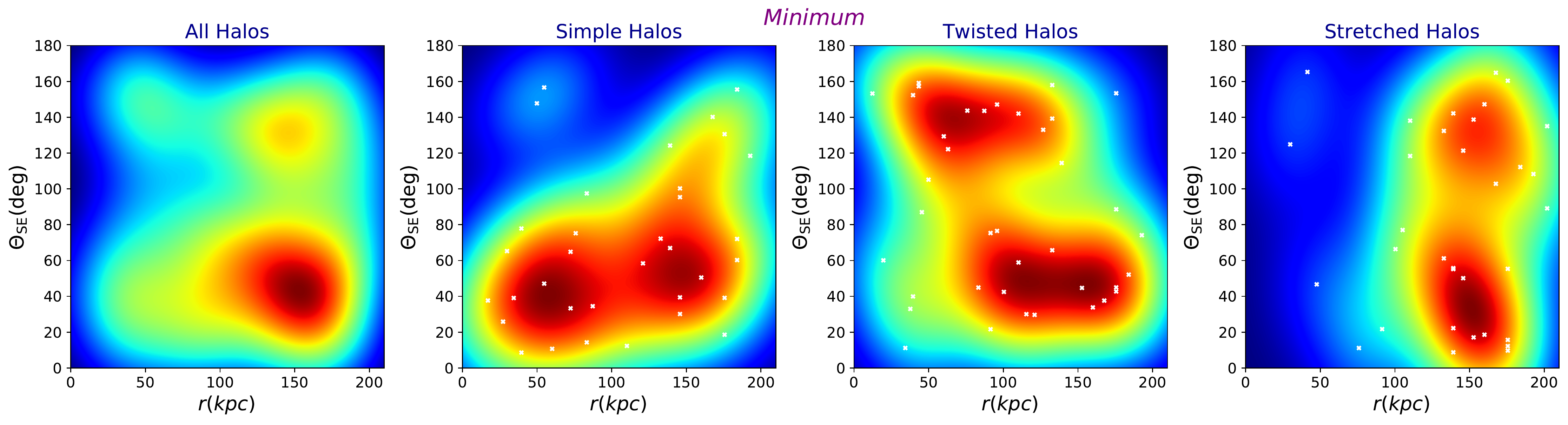}
\caption{2D distribution of the angle between the angular momentum of satellites and eigenvectors associated with the minimum eigenvalues of the inertia tensor located at the closest distance to individual satellites. Satellites of simple halos are in general more aligned with the minimum eigenvector. Twisted/Stretched halos show a bi-modal distribution 
of satellites. In each panel, marked crosses refers to individual satellites. 
\label{angle_satellites_min}}
\end{figure*}

\begin{figure*}
\center
\includegraphics[width=1.0\textwidth,trim =  5mm 2mm 2mm 1mm]
{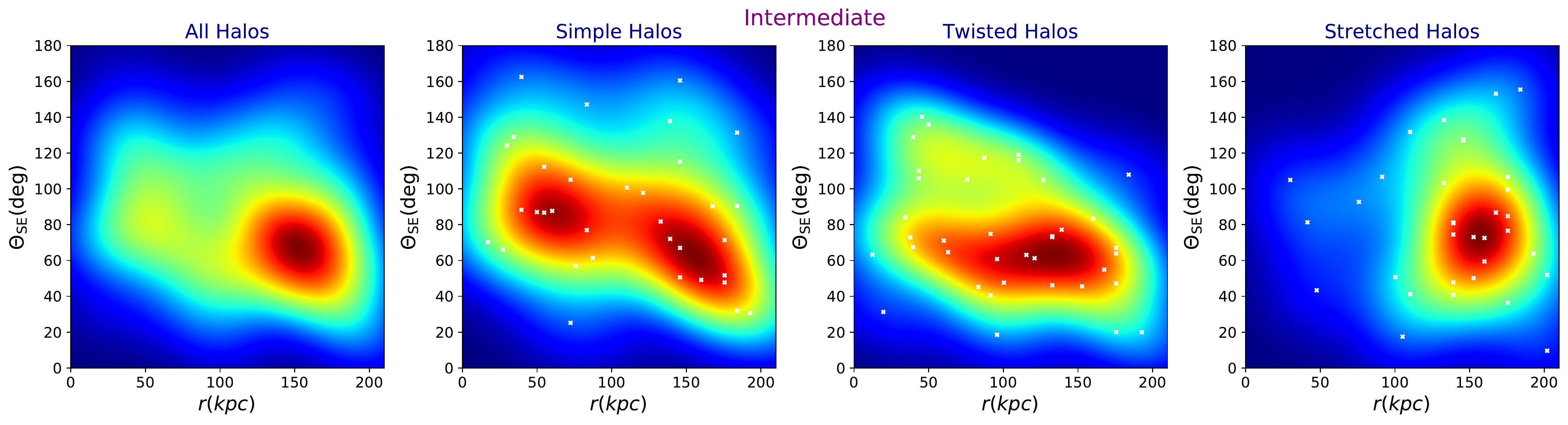}
\caption{2D distribution of the angle between the angular momentum of satellites and eigenvectors associated with the intermediate eigenvalues of the inertia tensor located at the closest distance to individual satellites. Satellites of simple halos are mostly orthogonal to the intermediate eigenvector. In each panel, marked crosses refers to individual satellites. 
\label{angle_satellites_inter}}
\end{figure*}

\begin{figure*}
\center
\includegraphics[width=1.0\textwidth,trim =  5mm 2mm 2mm 1mm]
{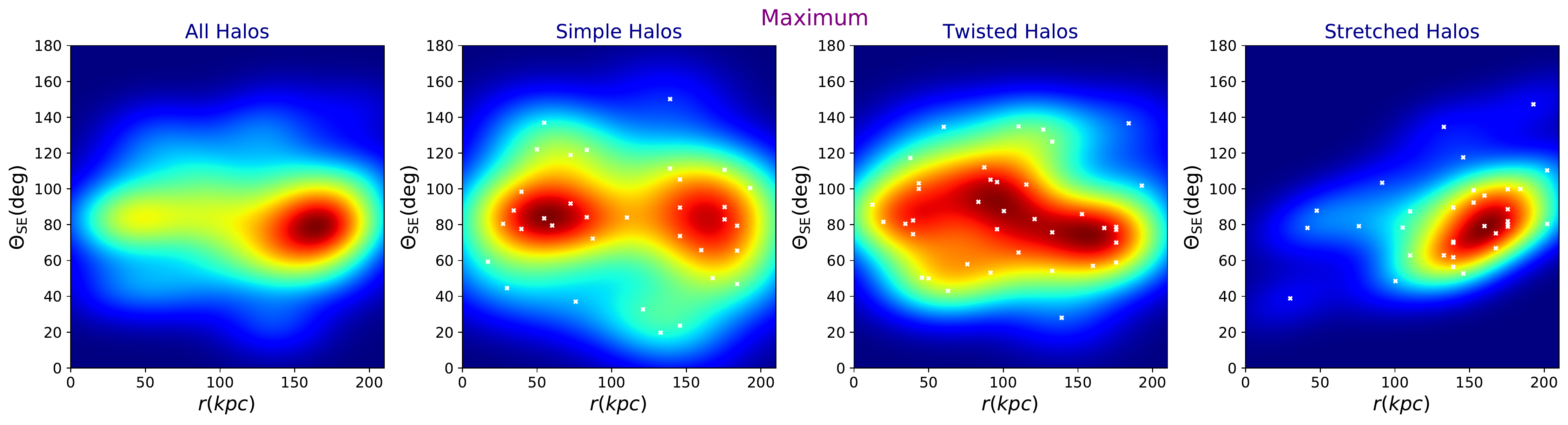}
\caption{2D distribution of the angle between the angular momentum of satellites and eigenvectors associated with the maximum eigenvalues of the inertia tensor located at the closest distance to individual satellites. Satellites of simple halos are almost orthogonal to the maximum eigenvector. In each panel, marked crosses refers to individual satellites. 
\label{angle_satellites_max}}
\end{figure*}

In Figures \ref{angle_satellites_min}-\ref{angle_satellites_max} we study the 2D distribution of the angle between the angular momentum of satellites and the closest eigenvectors associated with the minimum-to-maximum eigenvalues. Also to compute and track the angle profiles, we propose for the angles to be initially less than 90 \rm{deg}.
In each figure, from the left to right, we draw the 2D distribution of all, simple, twisted and stretched halos, respectively. 
There are few interesting take aways points from the above analysis:

$\bullet$ In Simple halos, the angular momentum of satellites is more aligned with the minimum eigenvector than the other two classes.

$\bullet$ In twisted and stretched halos, there is a bi-modality in the distribution of the angular momentum of satellites and the minimum eigenvectors computed at different locations. Where almost a half of satellites are anti-aligned (with the angles around 180 \rm{deg}) with the minimum eigen-vector, while the rest of them are in a similar range of angles to those associated with simple halos.

$\bullet$ There is also a bi-modality in the radial distribution of the satellites of simple and twisted halos. In particular, almost half of them are located closer than 100 \rm{kpc}, while the rest are between 100 to 200 \rm{kpc}. 

$\bullet$ While the angle profiles of the angular momentum of satellites and the minimum eigenvectors are peaked at around 40 \rm{deg}, the radial profile of all of angular momentum of satellites with the intermediate and maximum eigenvectors peak at around 70 \rm{deg} and 80 \rm{deg}, respectively. This means that satellites are generally in more aligned with the minimum eigenvector than the intermediate and maximum ones.

\begin{figure*}
\center
\includegraphics[width=1.\textwidth,trim =  5mm 2mm 2mm 1mm]
{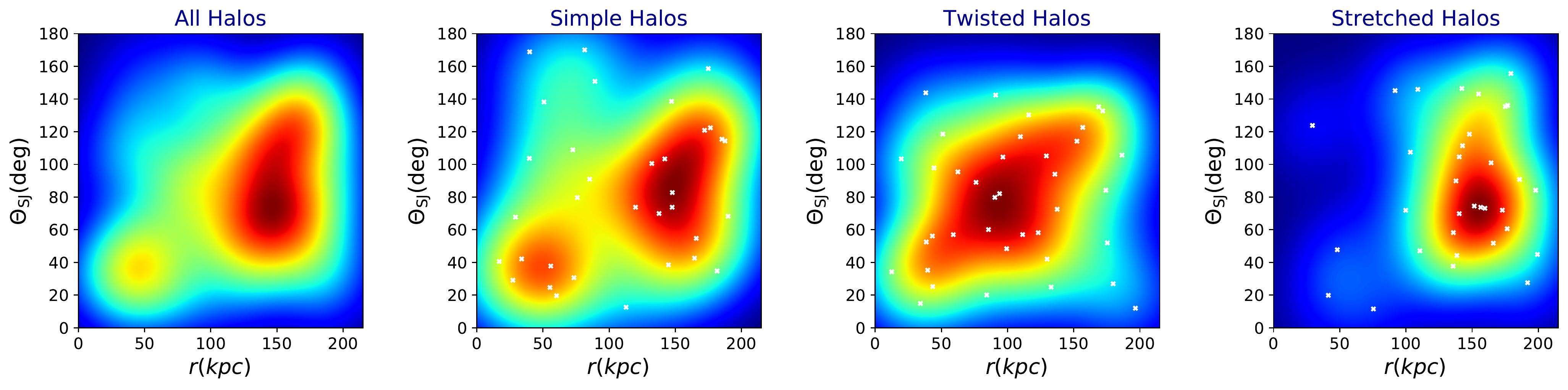}
\caption{2D distribution of the angle between the angular momentum of satellites and net specific angular momentum.
In  simple and twisted halos,  $\theta_{\mathrm{SJ}}$ grows with increasing the radius, while in the stretched halos, $\theta_{\mathrm{SJ}}$ has a flat distribution. In each panel, marked crosses refers to individual satellites.
\label{angle-satellites-jnet}}
\end{figure*}

Next, in Figure \ref{angle-satellites-jnet} we study the 2D distribution of the angle between the angular momentum of satellites and $\mathbf{j}_{\rm{net}}$. From the left to right, we present all, simple, twisted and stretched halos, respectively. 

From the plot, it is inferred that in the simple and twisted halos,  $\theta_{\mathrm{SJ}}$ mostly grow from left to right where closer by satellites are being more aligned with the $\mathbf{j}_{\rm{net}}$ than those farther out. There are however few cases where simple/twisted halos are anti-aligned with the $\mathbf{j}_{\rm{net}}$. On the contrary, in the stretched halos, $\theta_{\mathrm{SJ}}$ seems to have a flat distribution.

\begin{figure*}
\center
\includegraphics[width=1.0\textwidth,trim = 5mm 1mm 0mm 1mm]
{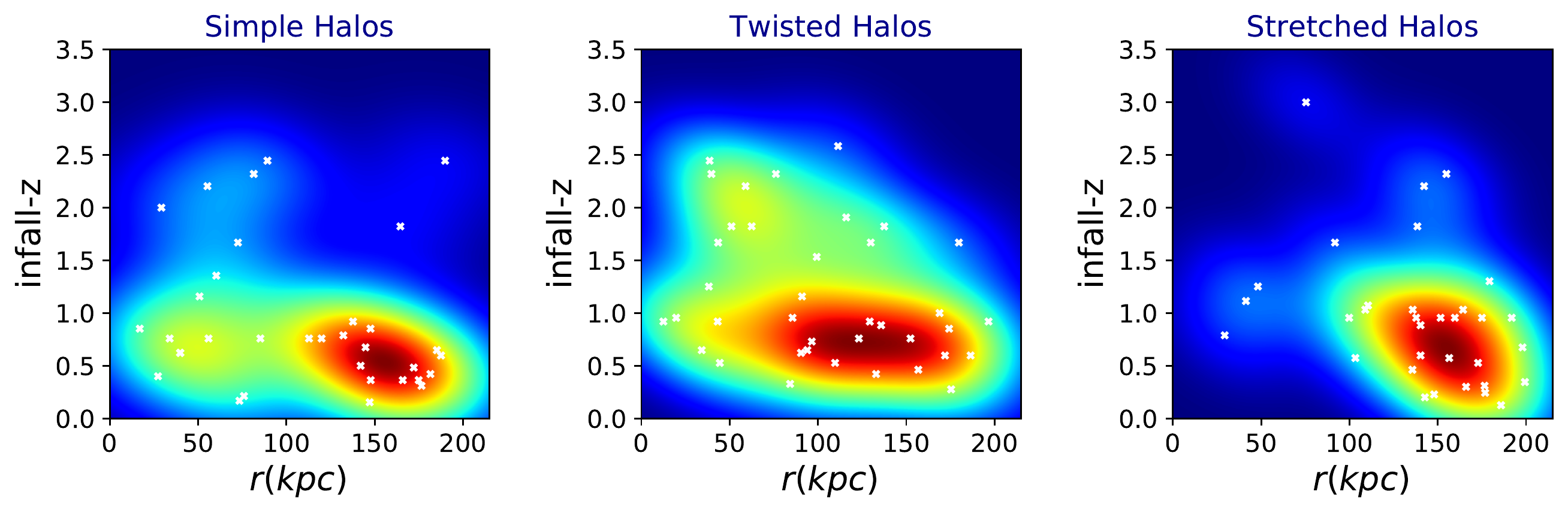}
\caption{ The 2D distribution of the infall time of satellites in different halo types in our sample.We have taken only satellites with masses above $10^7 M_{\odot}$ and at the distance of less than 200 \rm{kpc}. In each panel, marked crosses refers to individual satellites. 
\label{Infall-Sat-2D}}
\end{figure*}

\begin{figure}
\center
\includegraphics[width= 0.45\textwidth,trim = 5mm 1mm 0mm 1mm]
{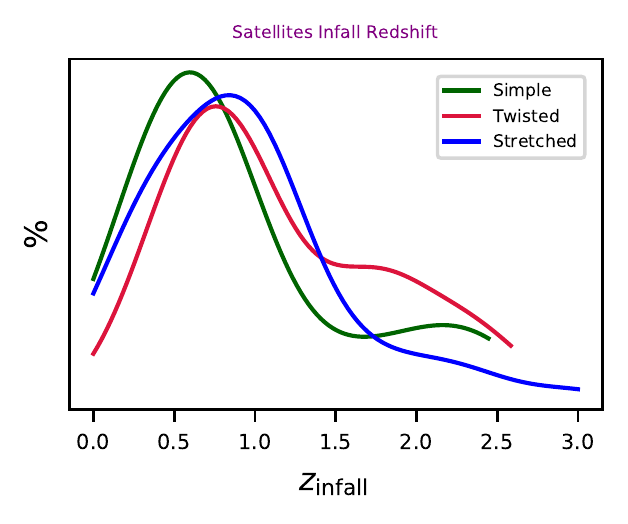}
\caption{ 1D distribution of the infall time for satellites in different halo types in our galaxy samples.
\label{Infall-Sat-1D}}
\end{figure}

Finally, in Figures \ref{Infall-Sat-2D}-\ref{Infall-Sat-1D}, we present the 2D and 1D distribution of the infall redshift vs the radius and infall redshift for satellites of different halo types. To infer the satellite's infall-time, we trace each of them backward in time down to the redshift where prior to this, the satellite does not belong to the central halo. 

From Figure \ref{Infall-Sat-2D}, it is inferred that, for simple and stretched halos, the infall-time is mostly between z = 0.3-1.0 and less than 25\% of satellites being accreted before $z > 1.0$. However, in twisted halos, there are some bi-modalities in the distribution of the infall-time of satellites with about 39\% of them being accreted at $z> 1.0$. Such bi-modality is seen as the little bump in the 1D distribution of the satellite's infall-time from Figure \ref{Infall-Sat-1D}. Again we should note that since the current sample is limited, care should be taken in any statistical conclusions!

\section{Connection to observations}
\label{Observation}

Thus far we assessed the shape of DM halo in MW like galaxies in TNG50 using different techniques. Below we connect these theoretical outcomes to recent observational constraints on these shapes. Furthermore, since the radial distribution of shape parameters are rather different between different techniques, we may be able to distinguish between them as well. 

Observationally, one may indicate the shape of DM halo without directly measuring this in MW galaxy. 
Below, we use the observational constraints on the shape of DM halos as provided in \cite{2016ARA&A..54..529B} and references therein.

One possible way to determine the DM halo shape in the MW is using the orbit of Sgr dwarf across the sky. It was shown that the geometry of the stream across the sky confirms an oblate to near spherical DM halo \citep{2001ApJ...551..294I,2005ApJ...619..800J}. More specifically, \cite{2005ApJ...619..800J} used radial velocities of a sample of (few)-hundred M giant candidates from Two Micron All Sky Survey (2MASS) catalog to trace streams of tidal debris associated with Sagittarius dwarf spheroidal galaxy (Sgr) which entirely encircle MW Galaxy. They strongly favoured an iso-density $ 0.83 \leq s \leq 0.92 $ at the distance between 13-60 \rm{kpc}
and ruled out $s \leq 0.7$ and $s \geq 1.1$ at 3$\sigma$. It is important to note that while for triaxial halos, both of shape parameters $s,q$ are by definition less than unity, in the context of axisymmetric halos, where the density profile is a function of $r^2 = (x^2 + y^2 + z^2/q^2)$, the flattening parameter $s$ can be smaller/larger than unity for the oblate/prolate halos. Therefore, we make a transformation between axisymmetric parameter (hereafter $q_{\mathrm{axisym}}$) and our triaxial based results. Below, we make the following transformation. If the $q_{\mathrm{axisym}}$ is larger than one, we read $s = q = q_{\mathrm{axisym}}$ and if the 
$q_{\mathrm{axisym}}$ is less than one, we shall take $s = q_{\mathrm{axisym}}$  while $q = 1$.

Later, \cite{2004ApJ...610L..97H} used line-of-sight velocities and compared the kinematics of \cite{2005ApJ...619..800J} M giant sample to the models of Sgr dwarf debris and showed that a portion of mapped trailing stream is dynamically young and thus does not imply a strong constraint of the shape. On the other hand, the leading stream consists of older debris and its dynamics provides a strong indication towards a prolate halo shape with $s = 5/3$. 

\cite{2007A&A...469..511K} used data from Leiden-Argentine-Bonn all sky 21-cm line survey and derived the 3D $\rm{H _I}$ density distribution for MW to constrain the galactic mass distribution.  They found a majority of DM particles can be modeled using an isothermal disk within $r \leq 40$ \rm{kpc}. Though the confirmation of DM disk is a hint for an oblate shape, they showed that a halo with a constant $s$ does not quite match with the observations and that the halo shape should be progressively prolate at larger distances. 

\cite{2010ApJ...712..260K} combined SDSS photometry, USNO-B astrometry and SDSS/Calar Alto spectroscopy and constructed an empirical 6D phase-space map of GD-1 stream of stars located at 15 \rm{kpc} from the galactic center and is believed to be debris from a tidally disrupted star cluster. Using an axisymmetric potential, made of stellar disk and DM halo, they found $s_{\Phi} \geq 0.89$, 
at the Galactocentric radii near to 15 \rm{kpc};
where $s_{\Phi}$ refers to the flattening in DM gravitational potential. 

\cite{2012MNRAS.425.1445G} used the kinematic and position data for $\sim 2000$ K dwarf stars located near the sun in a distance less than 1.1 \rm{kpc} from the galactic plane from \citep{1989MNRAS.239..605K} and determined the DM halo density in the MW. They reported a mild tension with the assumption of a spherical halo but consistent with an oblate DM halo with $s \geq 0.7$ or a local disc or a spherical DM halo with larger normalization.

\cite{2015ApJ...803...80K} used the tidal streams from Palomar5 (Pal5), the faintest and most extended, globular cluster in the MW. They estimated the DM halo shape to be nearly spherical with a  potential flattening $s_{\Phi} = 0.95^{+0.16}_{-0.12}$ at the  
heliocentric distance 23.6 \rm{kpc}. 
Using the proper motion of 75 globular clusters in Gaia DR2, \cite{2019A&A...621A..56P} estimated the mass and axis ratio of DM within $r \leq 20$ \rm{kpc}. They reported a prolated DM halo with $q_{\mathrm{axisym}} = 1.3 \pm 0.25$. This rules out very oblate DM halo with $q_{\mathrm{axisym}} < 0.8$ and very prolated halo with $q_{\mathrm{axisym}} > 1.9$ at 3$\sigma$. 

\begin{figure}
\center
\includegraphics[width=0.45\textwidth, trim = 7mm 2mm 2mm -2mm]{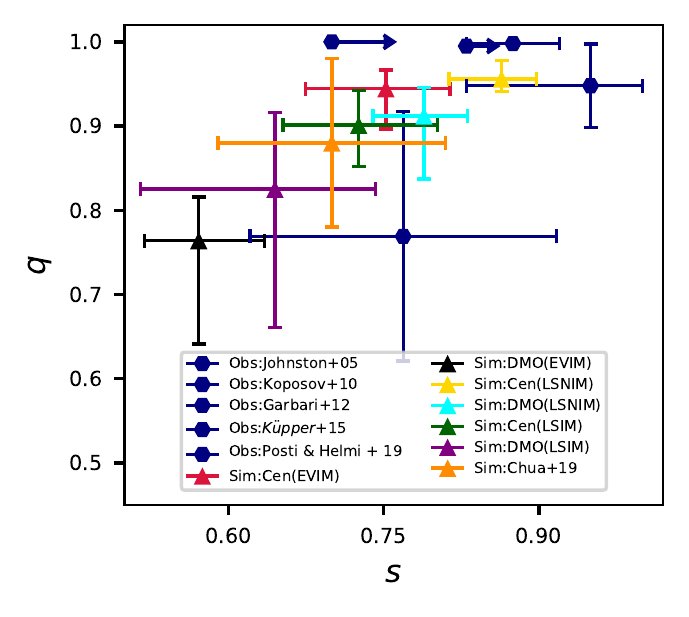}
\caption{Different observational constraints on the shape parameters of the DM halo. 
Overlaid on the plot are the median of $s,q$ for both of 
the central (Cen) and DMO simulations analyzed by our three methods as \rm{EVIM}, \rm{LSNIM} and \rm{LSIM}. Care must be taken that \cite{2010ApJ...712..260K} and  K\"{u}pper et al. +15 results are based on the shape of the potential and not the shape of the DM distribution.
\label{sq-DM-halo}}
\end{figure}

Figure \ref{sq-DM-halo} summarizes the above constraints on the shape parameters of the DM halo. 
Overlaid on the plot are the medians of \rm{EVIM},  \rm{LSNIM} and \rm{LSIM} for both of Cen and DMO simulations. 
We shall emphasize that all of the y-axes are located at $q = 1$, the slight displacement is for the clarity of the presentation. Out results are comparable with most of these results. The level of the agreement between the simulations ans observational results are indeed excellent.

\section{Summary and conclusion}
\label{conc}

In this paper, we used the hydrodynamic simulation of TNG50 and extracted a sample of 25 MW like galaxies identified using two different criteria. 
The first criterion is that the DM halo is belong to a mass range of $10^{12} M_{\odot}$ to $1.6 \times 10^{12} M_{\odot} $. The second is that the disk-frac, defined as the fraction of number of stars with orbital circularity parameter, $\varepsilon \geq 0.7$, is above 40\%. 
We computed the radial profile of shape parameters $(s,q)$ for DM halo. We exploited three different approaches in our shape analysis. In the first approach, we inferred the halo shape using an enclosed volume iterative method, \rm{EVIM}. In the second approach, we computed the shape using a local shell non-iterative method, \rm{LSNIM}. 
Finally, in the third approach, we calculated the shape using a local in shell iterative method, \rm{LSIM}.

The radial profile of the $s$ is fairly similar between \rm{EVIM}, \rm{LSIM} in the inner part of the halo but slightly deviates after the radius of 80 \rm{kpc}. On the contrary,, \rm{LSNIM} predicts a larger profile for $s$ parameter which is progressively diminishing from the inner to the outer part of the halo. On the other hand, the radial profile of the $q$ is very close between \rm{EVIM}, \rm{LSIM} up to 10 \rm{kpc} with a switch over behavior at larger radii where 
\rm{EVIM} gets closer to the \rm{LSNIM} than \rm{LSIM}.

Based on our shape analysis which are mainly taken from \rm{EVIM}, we classify DM halos in our galaxy sample into 3 main categories.
Simple halos develop well separated eigenvalues that never cross each other (based on EVIM). 
But,owing to the local fluctuations, they could pass through each other in \rm{LSIM}. However, since our halo classification is based on the \rm{EVIM}, name this class as simple halos. Furthermore, the eigenvector associated with the minimum eigenvalue in these halos is almost entirely parallel to $\mathbf{j}_{\rm{net}}$. There are in total 8 halos in this class.

Twisted halos establish some level of gradual rotation throughout their radial profile. The level of reorientation varies from one halo to the other but in general the halo is reoriented in this radial profile. There are in total 8 halos in this class of halos.

Stretched halos experience some levels of stretching (even in \rm{EVIM}) in their radial profiles, where different eigenvalues cross each other. Consequently, the angle of their corresponded eigenvectors with different vectors varies by 90 \rm{deg} at the location of stretching, thanks to the orthogonality of different eigenvectors. 
There are in total 9 different halos in this category.

We drew 3D ellipsoids for each category and established the rotation in the halo radial profile for each part.

We studied the main drivers of the DM halo shape. In this first study, we focused on three different drivers including the baryonic effects, impact of substructures. For the first driver, we computed the halo shape in dark matter only (DMO) simulations. We measured a smoother radial profile for $(s,q)$ with a triaxial/prolate halo shape. This means that baryonic effects tend to decrease the diskyness of halos. Remarkably, in DMO simulation, $s$ and $q$ are increasing with the radius. Accordingly, $T$ is decreasing for both approaches. Our analysis showed that halos are more simple in DMO than the full hydro-simulation. This is suggestive and indicates that twisted/stretched halos may have some baryonic reasons. Work is in progress to study these features in more details.

We examined the effect of substructures in the shape in both of full hydro-simulation and DMO simulations. Our analysis shows that in most cases the impact of substructures in the shape are subdominant in the inner part of the halo.

Furthermore, we also studied the location and angular momentum of MW satellites in the mass range $ M_{\star} \geq 10^{7} M_{\odot}$ and at the distance less than 200 \rm{kpc}. We computed the radial profile of the angle between the angular momentum of satellites and min,inter, max eigenvectors. Our analysis show that the satellites of simple halos are more aligned with the minimum eigenvector than the other two classes. Furthermore, in twisted and stretched halos, instead, there is a bi-modality in the distribution of the angular momentum of the satellites and the closest minimum eigenvector. 

Furthermore, the distribution of the infall time of satellites shows that the majority of them is accreted onto the main halo between $z = 0.3-1.0$.

Finally, we connected our theoretical predictions for shape parameters to some of the recent  observational studies. We overlaid our theoretical results on the top of few well studied tracers such as stream of tidal debris, GD-1 stream of stars, K dwarf stars and globular clusters from Gaia DR2 and found a fairly good agreement. In a companion paper, we aim to make a comprehensive study of the remaining possible drivers in the halo shapes. 

Since the radial profile of the shape parameters are not the same, we may hope that more detailed observations at different radii distinguish between different methods. For instance, it would be fascinating to use the spectroscopic data from Hectochelle in the Halo at High Resolution (H3)
survey \citep{2019ApJ...883..107C} and compute the radial dependence of the shape for DM halo.

Throughout this work, we only studied the shape of the DM halo. 
In \cite{2020arXiv201212284E}, we generalize this study to the case of Stellar halo and their possible connection to that of DM shape as studied here.

\begin{figure*}
\center
\includegraphics[width=0.7\textwidth, ,trim = 6mm 1mm 2mm 1mm]{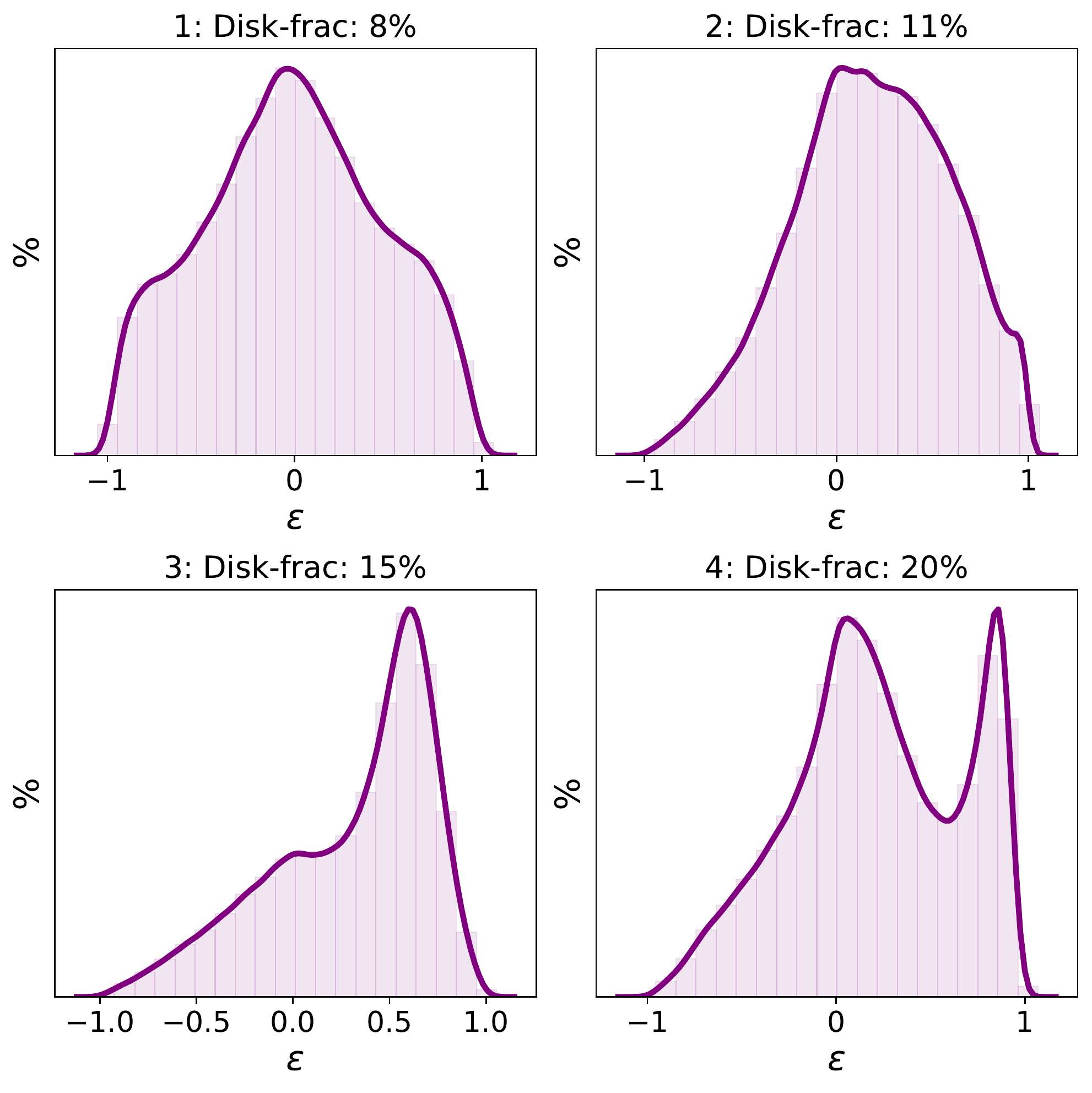}
\caption{Distribution of orbital circularity parameter for a sample of 4 galaxies with mass similar to MW galaxy. On the top of each plot, we present the Disk-frac as a reference. \label{Non-Disky1}}
\end{figure*}

\section*{Data Availability}
Data directly related to this publication and its figures is available on request from the corresponding author. The IllustrisTNG simulations themselves are publicly available and accessible at \url{www.tng-project.org/data} \citep{2019ComAC...6....2N}, where the TNG50 simulation will also be made public in the future.

\section*{acknowledgement}
It is a great pleasure to thank David Barnes, Angus Beane, Ana Bonaca, Dylan Nelson, Sandro Tacchella, Matthew Smith, and Annalisa Pillepich for the very insightful conversations. We especially acknowledge Charlie Conroy for his assistance with making the connection to observations. We thank the referee for their constructive comments that improved the quality of this paper. R.E. acknowledges the support by the Institute for Theory and Computation at the Center for Astrophysics. We thank the supercomputer facility at Harvard where most of the simulation work was done. MV acknowledges support through an MIT RSC award, a Kavli Research Investment Fund, NASA ATP grant NNX17AG29G, and NSF grants AST-1814053, AST-1814259 and AST-1909831. SB is supported by Harvard University through the ITC Fellowship. FM acknowledges support through the Program "Rita Levi Montalcini" of the Italian MIUR. The TNG50 simulation was realized with compute time granted by the Gauss Centre for Supercomputing (GCS) under GCS Large-Scale Projects GCS-DWAR on the GCS share of the supercomputer Hazel Hen at the High Performance Computing Center Stuttgart (HLRS).

\textit{Software:} matplotlib \citep{2007CSE.....9...90H}, numpy \citep{2011CSE....13b..22V}, scipy \citep{2007CSE.....9c..10O}, seaborn \citep{2020zndo...3629446W}, pandas \citep{mckinney2010data}, h5py \citep{2016arXiv160804904D}.

\appendix 

\section{Non-Disky galaxies}
\label{non-disk}
As the main focused of this paper, so far we merely studied the disky MW like galaxies with more than 40\% of stars living in the disk, defined with $\varepsilon \geq 0.7$. From Figure \ref{epsilon-TNG50} we inferred a very similar distribution for the orbital circularity parameter.  
In this appendix, we proceed further and study the distribution of the orbital circularity parameter for a sub-sample of galaxies with less fraction of stars living in the disk. In Figure \ref{Non-Disky1} we present the distribution of $\varepsilon$ for 4 galaxies with the mass similar to the MW galaxy. On the top of each plot, we labeled the Disk-frac. The plot considers 4 different Disk-fracs: $8\%, 11\%, 15\%, 20\%$. 

Unlike the case of disky MW like galaxies with very similar distribution, here the distribution of $\varepsilon$ varies a lot between different galaxies. It will be interesting to compute the shape for these non-disky galaxies. This analysis is however beyond the scope of this work and is left to a future study. 

\section{Convergence Check for shape analysis}
\label{Convergence}

As already mentioned in Sec. \ref{EVIM-Shape}, our first shape finder algorithm, \rm{EVIM}, relies on an iterative approach in which the algorithm is stopped after the shape parameters $(s,q)$ are converged such that
\rm{max}$(T_s, T_q ) \leq  10^{-3}$ where 
$ T_s \equiv \left((s-s_{\rm{old}})/s\right)^2$ and $ T_q \equiv \left((q-q_{\rm{old}})/q\right)^2 $ 
refer to the residual of the shape parameters $(s,q)$. Here we describe in more detail how the convergence is established in our approach. Furthermore, as a sanity check, we will also present few examples with progressive reduction of the residuals from the above algorithm. 

The convergence check in our shape finder algorithm proceed as follows. First, we leave the system for at  least 12 iterations before proposing any convergence criteria. This allows the system to get stabilized since its first couple of iterations are required to get deviated from the initially proposed spherical symmetry. Then, to make sure that the convergence has truly established, we do not terminate the iteration once we get below the above threshold but allow the system to proceed and wait for having at least 15 consecutive  cases with the residual below $10^{-3}$. Since at every iteration the spheroid is getting deformed, it may well happen that the system experiences some sudden jumps and the residual increases from one iteration to the other. If that occurs, we will set the counter to zero and again seek for 15 consecutive cases below the above threshold. This is very ensuring condition and in most cases reduce the residual substantially below $10^{-3}$. 
\section{Halo Classification}
\label{classification}

As already mentioned in the main text, based on our shape analysis, we can identify  
different halo types; the so called simple, twisted and stretched halos. Here we present the radial profile of the Axes/r ratio, the angle of min-inter-max eigenvectors with
the unit vectors along the x, y, z directions of the simulation box
and the shape profile for all of 25 halos in our galaxy sample.

Furthermore, in order to have unambiguous association of angles at the initial point and avoid the possibility of the exchange of angles from very small to close to 180, we propose that all of the angles are initially less than 90 \rm{deg}. Depending on the halo type, some of these angles could grow and get bigger than 90 \rm{deg} throughout their radial profiles. This is done by simply making a mask over the angle associated with the initial location and apply this to the entire radial profile of the angles.

\subsection{Profile of simple/Stretched halos }
\label{classification_1}
We start with the simple halo class.
In Figures \ref{Simple1}-\ref{Simple2}, we present the radial profile of Axes/r, angles and shape parameters for this type. Overlaid on each figure, we also present the results from the \rm{LSIM} method. Quite remarkably, the results of \rm{EVIM} for the Axes/r and shape parameters are fairly close to that of \rm{LSIM}. However, since \rm{LSIM} is sensitive to the local details of the shape, it is seen that there are some extra crossing of lines from \rm{LSIM} which leads to some extra rotations in \rm{LSIM}. Because of this, in \rm{LSIM}, we name this class as simple/stretching. However, as we take \rm{EVIM} as the main method, to simplify the classification, we call this group as simple halos. The results of \rm{EVIM} explicitly show that  the level of halo rotation is very minimal. There are in total 8 halos in this category.

\begin{figure*}
\center
\includegraphics[width=1.02\textwidth,trim = 6mm 1mm 2mm 1mm]{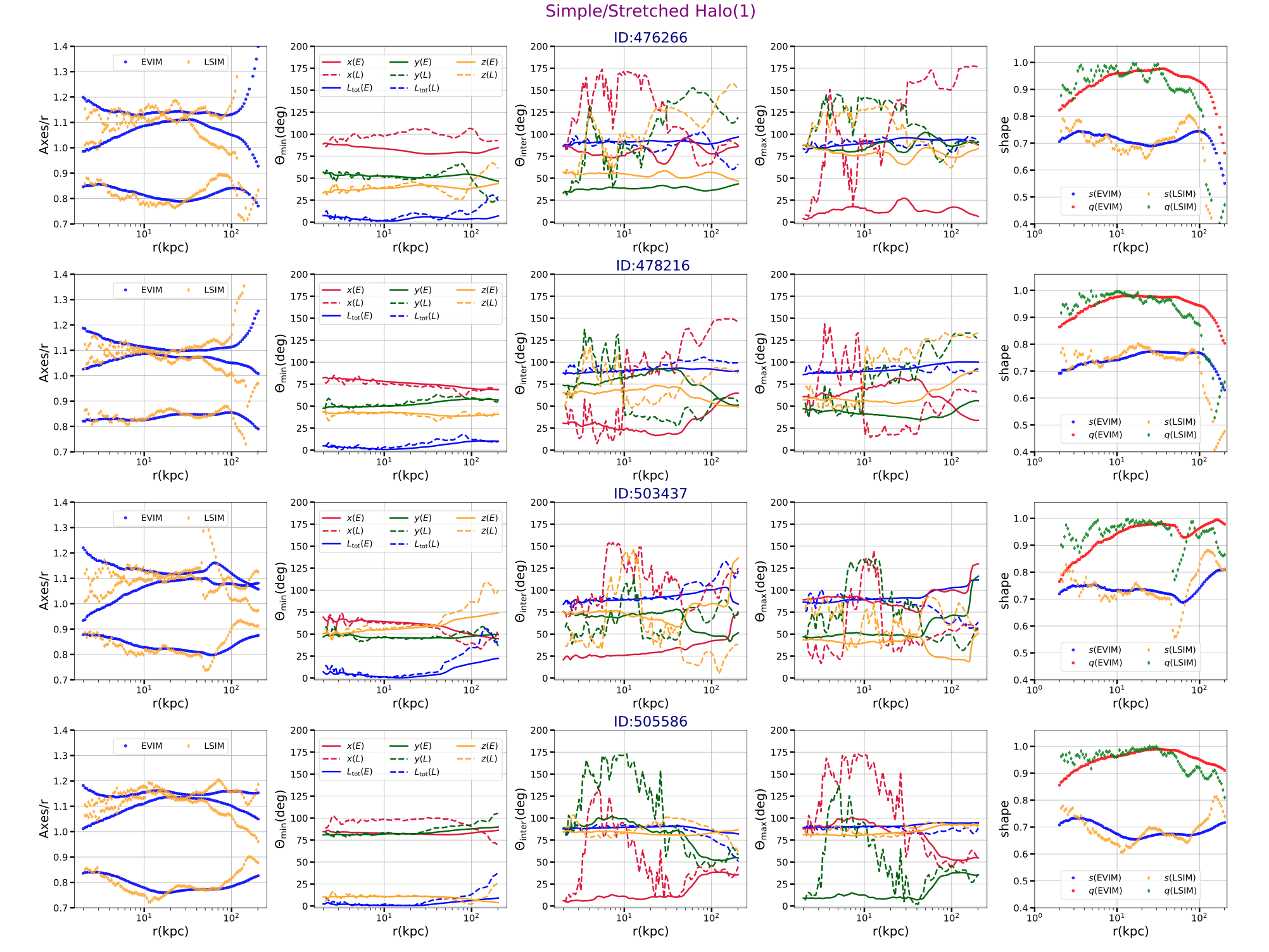}
\caption{ The radial profile of Axes/r, angles and shape parameters for the simple/Stretched halos. 
 \label{Simple1}}
\end{figure*}

\begin{figure*}
\center
\includegraphics[width=1.02\textwidth,trim = 6mm 1mm 2mm 1mm]{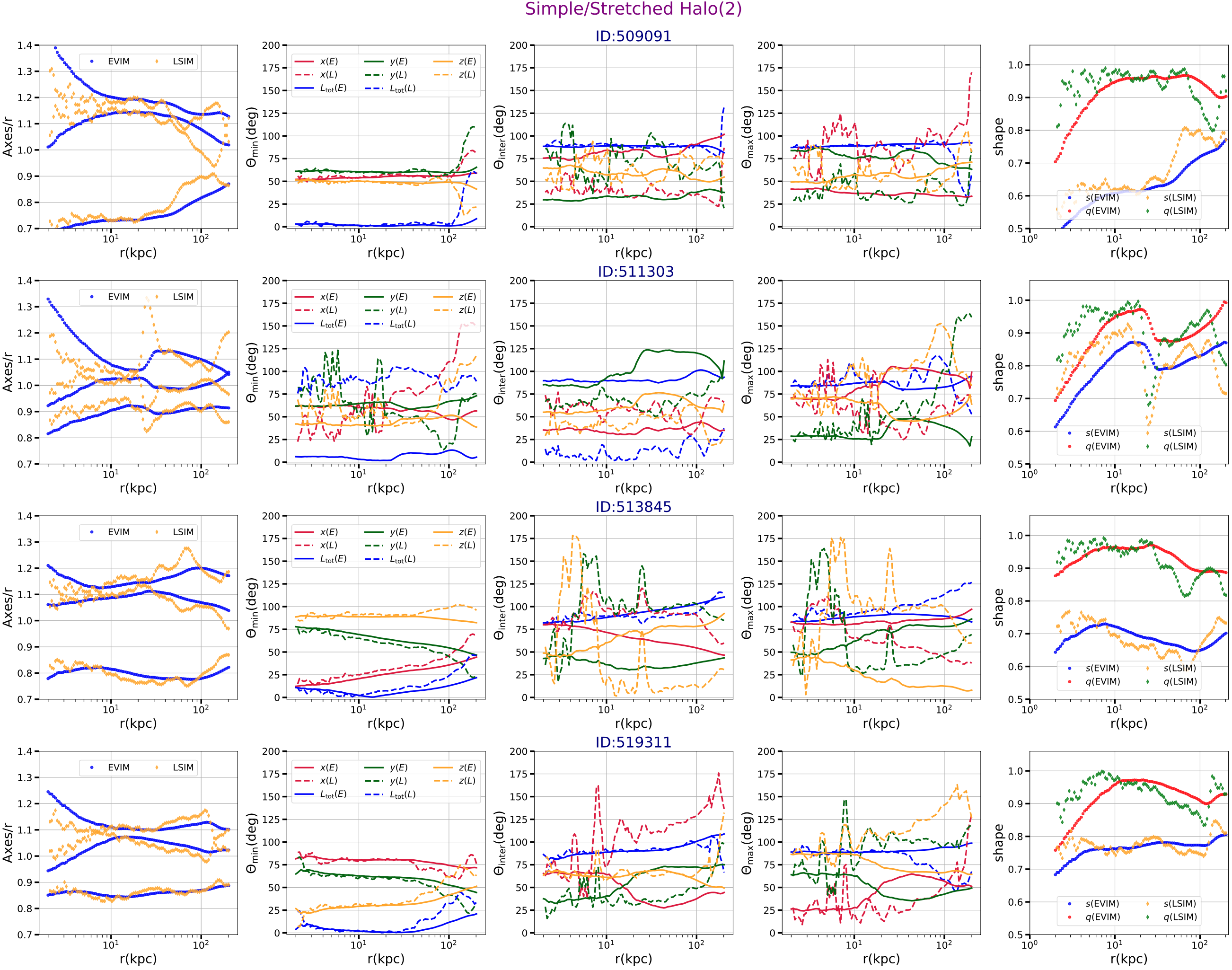}
\caption{ (Continued) The radial profile of Axes/r, angles and shape parameters for the simple/Stretched halos.
 \label{Simple2}}
\end{figure*}

\subsection{Profile of twisted halos }
\label{classification_2}
Next, we study the twisted halos. In Figures \ref{TwistedHalos1}-\ref{TwistedHalos2}, we analyse the radial profile of Axes/r, angles and shape parameters for this type. It is evident that halos in this category experience some gradual level of rotations from 50 to 100 \rm{deg}s in their radial profiles. There are in total 8 halos in this category.
Overlaid on each figure, we also present the results from the \rm{LSIM}. Interestingly, the outcome of \rm{EVIM} and \rm{LSIM} are fairly close to each other.

\begin{figure*}
\center
\includegraphics[width=1.02\textwidth,trim = 6mm 1mm 2mm 1mm]{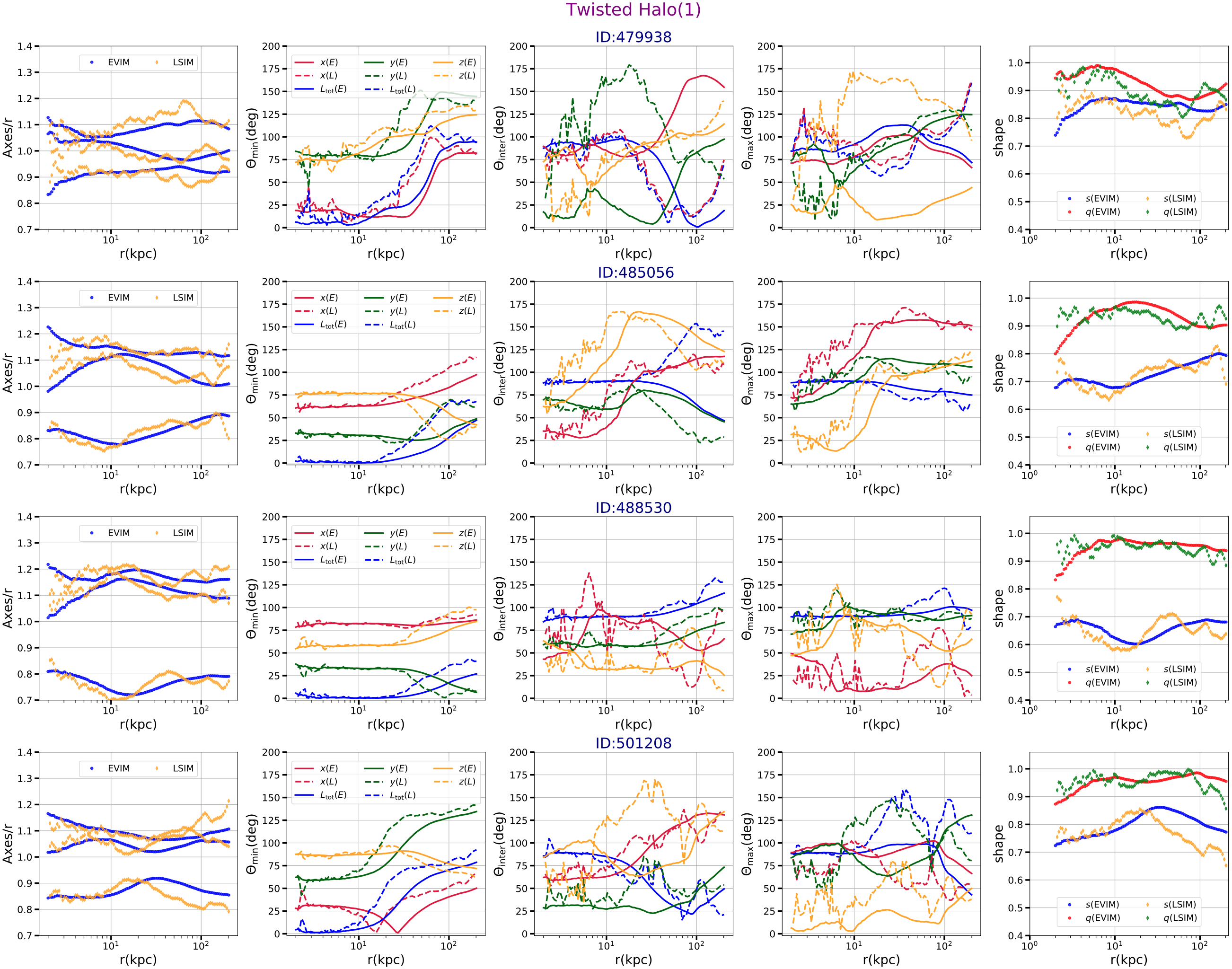}
\caption{ The radial profile of Axes/r, angles and shape parameters for the twisted halos. 
 \label{TwistedHalos1}}
\end{figure*}

\begin{figure*}
\center
\includegraphics[width=1.02\textwidth,trim = 6mm 1mm 2mm 1mm]{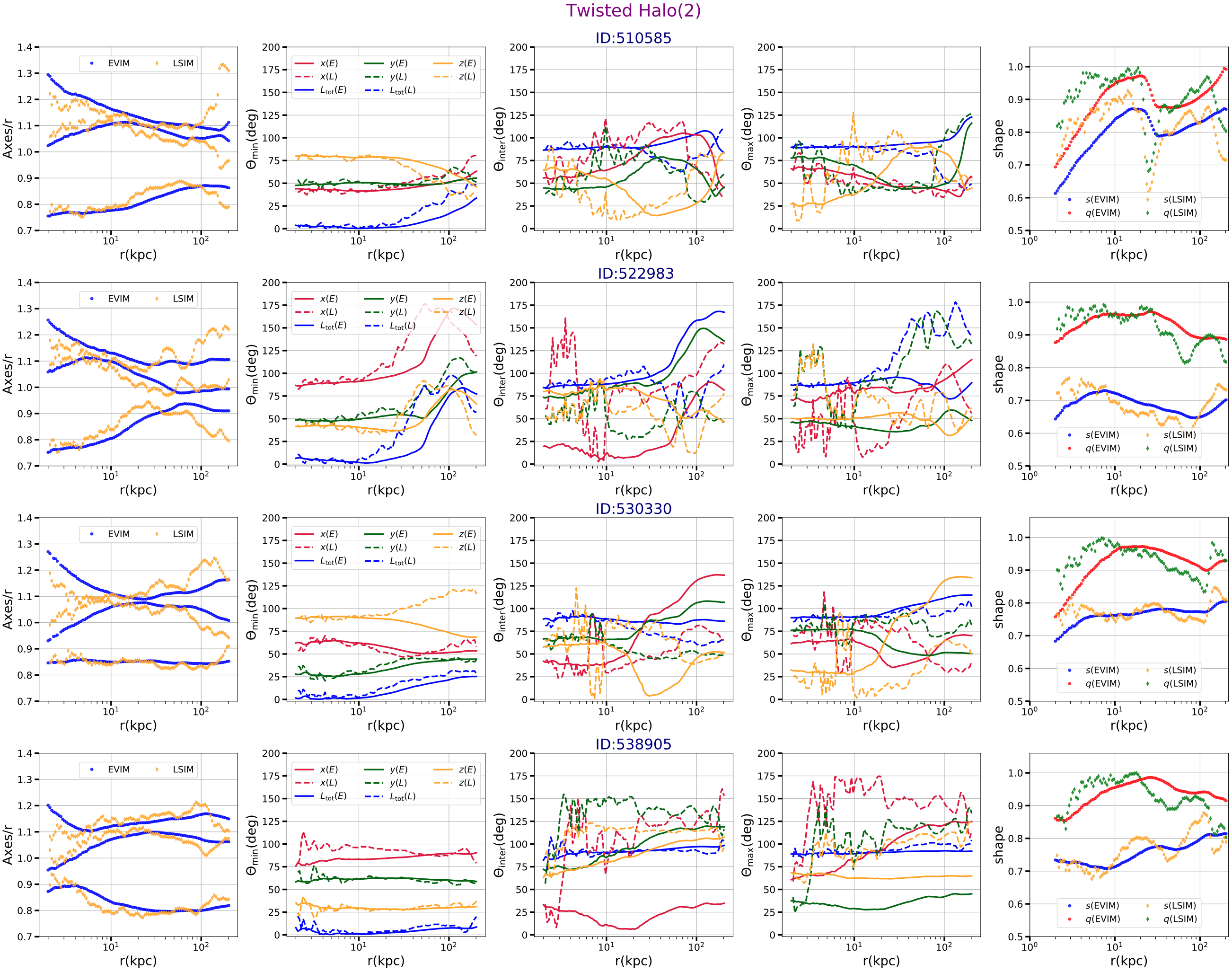}
\caption{ (Continued) The radial profile of Axes/r, angles and shape parameters for the twisted halos.
 \label{TwistedHalos2}}
\end{figure*}

\subsection{Profile of stretched halos}
\label{classification_3}

Finally, we analyse the stretched halos. In Figures \ref{Stretched1}-\ref{Stretched2}, we analyse the radial profile of Axes/r, angles and shape parameters for this type. Halos in this category, experience one (or more) stretching where different eigenvalues cross each other. Consequently, the halo experience a change of angle of order 90 \rm{deg} at the location of stretching. There are in total 9 halos in this class of halos. Overlaid on each figure, we also present the results from the \rm{LSIM}. Interestingly, the outcome of \rm{EVIM} and \rm{LSIM} are fairly close to each other.

\begin{figure*}
\center
\includegraphics[width=1.02\textwidth,trim = 6mm 1mm 2mm 1mm]{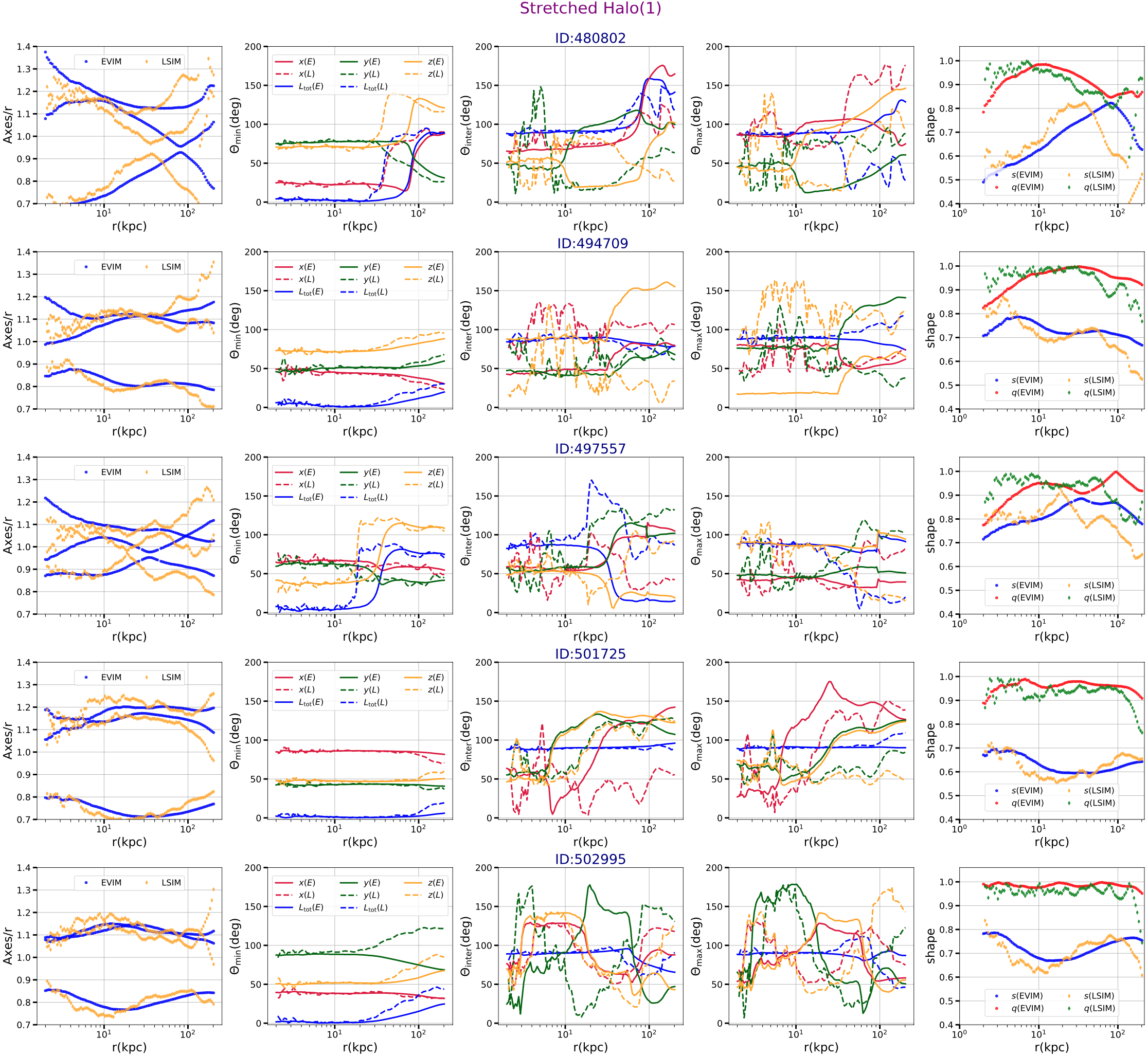}
\caption{ The radial profile of Axes/r, angles and shape parameters for the stretched halos.
 \label{Stretched1}}
\end{figure*}

\begin{figure*}
\center
\includegraphics[width=1.02\textwidth,trim = 6mm 1mm 2mm 1mm]{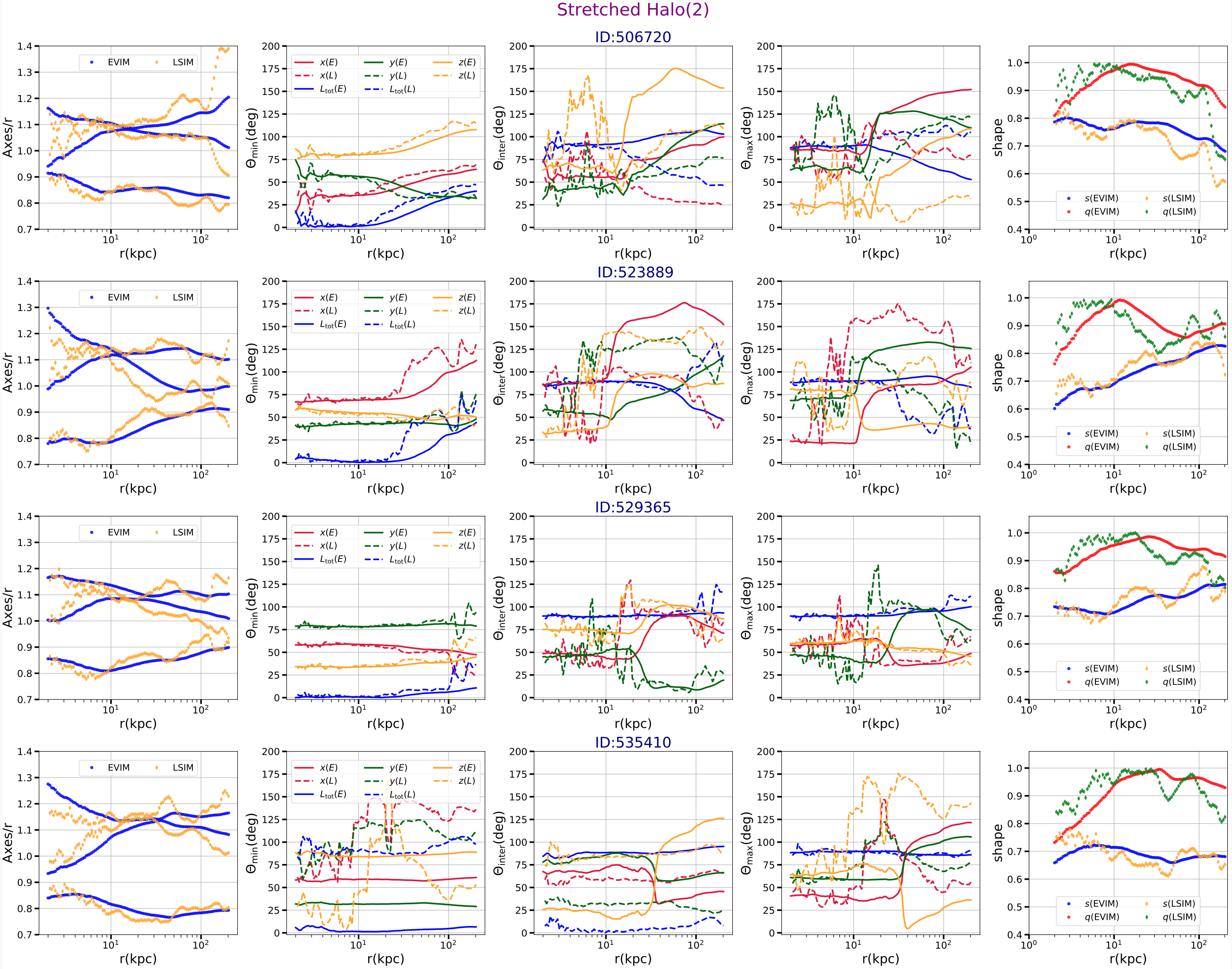}
\caption{ (Continued) The radial profile of Axes/r, angles and shape parameters for the stretched halos.
 \label{Stretched2}}
\end{figure*}

\section{Impact of weighting factor in the shape}
\label{Weighting-factor}
\begin{figure*}
\center
\includegraphics[width=0.9\textwidth,trim = 6mm 1mm 2mm 1mm]{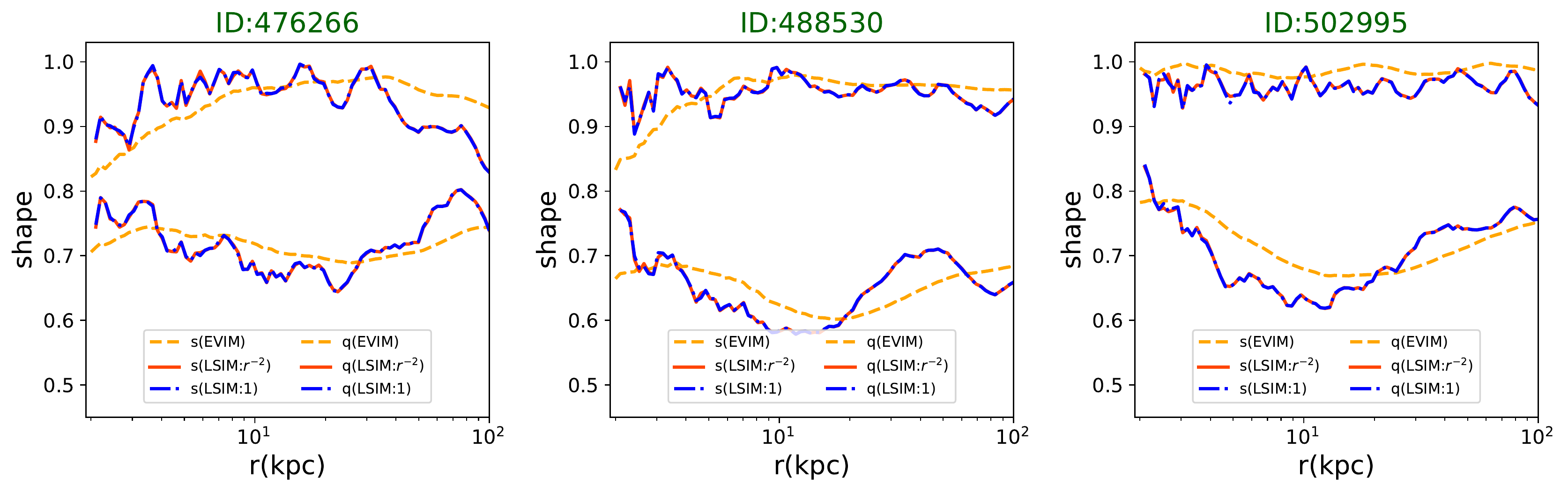}
\caption{ Impact of changing the weighting factor in shape parameters in a sub-sample of 3 MW like galaxies.
 \label{shape-method}}
\end{figure*}

Having presented different methods in analysing the DM halo shape, here we make a final comparison between these methods. Since the inertia tensor depends on the weighting factor, here we check the impact of different choices in the final shape. In Figure \ref{shape-method}, we examine the impact of $1/r^2$ and unity weighting factors in LSIM in the shape parameters. Overlaid on the figure, we also present the results from the EVIM. It is evident that the results of the LSIM from the above two choices of the weighting factors are almost the same. This is reasonable since we are dealing with very thin shells where the elliptical radii is almost one.  

\end{document}